\documentclass[12pt,onecolumn]{aastex6}
\renewcommand{\baselinestretch}{1.}
\usepackage{graphicx}	
\usepackage{amsmath}	
\usepackage{amssymb}	
\usepackage{longtable}
\usepackage{float}

\begin{document}
\title{Giant planets: good neighbors for habitable worlds?}
\author{Nikolaos Georgakarakos\altaffilmark{1}, Siegfried Eggl\altaffilmark{2}}
\and
\author{Ian Dobbs-Dixon\altaffilmark{1}}
\altaffiltext{1}{New York University Abu Dhabi, Saadiyat Island, Abu Dhabi, PO BOX 129188, UAE}
\altaffiltext{2}{California Institute of Technology, Jet Propulsion Laboratory, 4800 Oak Grove Drive, Pasadena, CA 91109, USA}

\begin{abstract}
The presence of giant planets influences potentially habitable worlds in numerous ways. Massive celestial 
neighbors can facilitate the formation of planetary cores and modify the influx of asteroids and comets towards Earth-analogs later on. 
Furthermore, giant planets can indirectly change the climate of terrestrial worlds by gravitationally altering their orbits. 
Investigating 147 well characterized exoplanetary systems known to date
that host a main sequence star and a giant planet we show that the presence of 'giant neighbors' 
can reduce a terrestrial planet's chances to remain habitable, even if both planets have stable orbits.  In a small fraction of systems, however, giant planets slightly increase the extent of habitable zones provided the terrestrial world has a high climate inertia.  
In providing constraints on where giant planets cease to affect the habitable zone size in a detrimental fashion,  
we identify prime targets in the search for habitable worlds. 
\end{abstract}


\section{Introduction} 

The discovery of planets of similar size to that of the Earth has led to
numerous interdisciplinary research activities about the potential habitability of other worlds (Cockell et al. 2016). 
Previous studies suggest that the presence of a giant planet in the same extrasolar planetary system can affect 
the gestation and evolution of potentially habitable planets in various ways.
During the planetary system's formation, a giant planet can operate as a dynamical barrier that blocks the inward migration of 
proto-planetary cores (Izidoro et al. 2015). This is important for the gestation of habitable planets, since the absence of a giant planet allows super-Earths to migrate inwards with the danger of 
strongly depleting the terrestrial planet forming zones of accretion material (Izidoro et al. 2014).  
Systems where hot Jupiters form in the outer parts of the disc and then migrate inwards seem to be unfavorable places for habitable worlds. Yet, it is not only possible to have terrestrial planets in the habitable zones\footnote{The habitable zone is the region around a star where a planet with an Earth-like
atmosphere on a circular orbit can support liquid water on its surface, see Kasting et al. (1993)}of such systems, but it is also likely that these planets have a substantial amount of liquid water due to the formation and inward migration history of the giant planet (Fogg {\&} Nelson 2007; Mandell et al. 2007). 
While the presence of a giant planet close to the habitable zone can mean dynamical chaos and instability for Earth-analogs (Sandor et al. 2007),
giant planets residing within the habitable zone can be hosts to habitable Trojan planets  (Schwarz et al. 2005) or habitable exomoons (Heller
et al. 2014), assuming the latter were not removed through an evection resonance during the planetary migration phase (Spalding et al. 2016).   
Even after the planetary formation phase giant planets continue to influence habitable conditions on terrestrial neighbors.  Large impacts of minor bodies on terrestrial planets, for instance, may be decisive for the existence of liquid water and the evolution of life (Kring 2003; Dvorak et al. 2012) and the presence of giant planets can increase the impact flux of comets and other minor bodies on terrestrial planets and hence the delivery of life enabling volatiles (Horner {\&} Jones 2008, 2009; Horner
et al. 2010; Grazier 2016).  Once life has emerged, however, impacts of asteroids and comets can trigger mass extinction events (Davis et al. 1984; Kaiho {\&} Oshima 2017).
The sometimes contradictory nature of the aforementioned processes shows that the role giant planets play for habitable worlds is not clear cut. 

This work is concerned with yet another vital piece in the puzzle. 
For planets such as our Earth the incoming stellar radiation constitutes the main energy source that 
determines the global climate on long timescales (Kasting et al. 1993). Consequently, the quantity and spectral distribution of insolation defines where a terrestrial planet can host liquid water close to its surface, which is the prime condition for habitability.
Planetary atmospheres can react strongly to changes in the incoming starlight, so that variations in some of the orbital elements or the obliquity of the planet constitute an important factor in assessing a planet's capacity to remain far from runaway states (Dressing
et al. 2010; Spiegel et al. 2010; Bolmont et al. 2016).
This is where giant planet 'neighbors' enter the picture.
The sheer presence of a giant planet in an exoplanetary system alters the orbit of a potentially habitable world over time. This effect may be very mild - a small change in the orbital elements of the terrestrial planet that results in spending some time 
outside the habitable insolation limits -  or very dramatic, leading to an eventual escape of the terrestrial planet from the system.  
In our Solar System for instance, the Earth's orbital eccentricity oscillates from 0 to 0.06 due to the long term perturbations of the other planets over a period of a few million years (Laskar et al. 2004). These so-called Milankovitch cycles have been linked to recent glaciation and interglacial periods (Horner {\&} Jones 2011;
Kaufmann {\&} Juselius 2016).
While the Earth experiences such variations over long timescales, this may not be the case for other terrestrial planets in exoplanetary systems. 
The dynamical interactions of planets in the habitable zone with a more massive planet have been extensively studied in the past (e.g. Sandor et al. 2007; Menou {\&} Tabachnik 2003; Jones et al. 2005; von Bloh et al. 2007; Kopparapu {\&} Barnes 2010; Dvorak et al. 2010; Carrera et al. 2016; Agnew
et al. 2017). Previous studies mainly focused on the orbital stability of the terrestrial planet assuming the habitable zone limits to be independent of the terrestrial planet's orbit. 
This, assumption, however, is only valid for terrestrial planets on a circular, unperturbed orbit. In this work, we study the effects of the presence of a giant planet on the habitability of an additional Earth-like planet in a more self-consistent manner. 
 By using analytical solutions for the evolution of a terrestrial planet under the gravitational perturbations of the giant planet we are able to calculate the actual insolation received by a potentially habitable world. 
Having estimates of the insolation extrema and variability\footnote{We refer to the extrema and variability during the long term (secular) orbital evolution of the terrestrial planet}for any given system configuration at our disposal, we are able to determine more realistic habitable zone limits. 

 \section{Habitable Zones and Insolation} \label{sec:met}
The practically circular orbit of the Earth around the sun and the relatively small variations in the luminosity of our star entail that we receive 
a near constant amount of sunlight on a permanent basis.  
For similar systems consisting of a star and a terrestrial planet on a fixed circular orbit, the limits of the (classical) habitable zone 
(CHZ) read (Kasting et al. 1993)
\begin{equation}
\label{eq:dist}
r_I=\left(\frac{L}{S_I}\right)^{\frac{1}{2}} \hspace{0.4cm}\mbox{and}\qquad r_O=\left(\frac{L}{S_O}\right)^{\frac{1}{2}},
\end{equation}
where $r$ is the distance of the planet to its host star in astronomical units, $L$ is the host star's luminosity in solar luminosities, and $S_I$ and $S_O$ 
are `spectral weights' corresponding to the number of solar constants  
that trigger a runaway greenhouse process (subscript $I$) evaporating surface oceans,
or a snowball state (subscript $O$) freezing oceans on a global scale. These spectral weights are functions of the effective temperature 
of the host star. As such they take the specific wavelength distribution of a star's light into account.
They constitute the most popular way to estimate the reaction of a terrestrial planet's climate to the amount and spectral
distribution of the incoming stellar radiation based on parametrization of
globally averaged radiative-convective energy balance models (Kasting et al. 1993; Kopparapu
et al. 2013, 2014). 
The latter provide `spectral weights' associated with the borders of the inner and outer edge of the habitable zone:
\begin{eqnarray}
\label{kop1}
S_I&=&1.107+1.332\cdot10^{-4}T_c+1.580\cdot10^{-8}T^2_c-8.308\cdot10^{-12}T^3_c-1.931\cdot10^{-15}T^4_c\\
\label{kop2}
S_O&=&0.356+6.171\cdot10^{-5}T_c+1.698\cdot10^{-9}T^2_c-3.198\cdot10^{-12}T^3_c-5.575\cdot10^{-16}T^4_c.
\end{eqnarray}
 $T_c=T_{eff}/1\;K-5780$, with $T_{eff}$ being the effective temperature of the star, while the value of 5780 used in the above fit formulae corresponds to the effective temperature $T_{\odot}$ of the Sun.
Although the coefficients in equations (2) and (3) refer to a planet of one Earth mass, they have also been evaluated for different terrestrial planet masses (Kopparapu
et al. 2014).
In order to calculate the habitable zone borders the luminosity of the host star has to be known.
In case stellar luminosities have not been observed directly, one can use stellar radii $R$ and effective temperatures $T_{eff}$ instead:
\begin{equation}
\label{eq:dist1}
\frac{L}{L_\odot}=\left(\frac{R}{R_{\odot}}\right)^2\left(\frac{T_{eff}}{T_{\odot}}\right)^4,\nonumber
\end{equation}
where $R_{\odot}$ is the Solar radius. The corresponding uncertainties are best determined via non-linear estimators.

\section{Climate Inertia}
Classical habitable zones borders that are calculated from spectral weights are based on the implicit assumption that the energy flux arriving at
the top of the atmosphere of a potentially habitable world remains constant on timescales long enough so that
incoming and outgoing radiation reach an equilibrium state.
Stable climatic conditions on the planet are linked to such equilibrium states.
The distance of a planet to its star and consequently the amount of light a planet receives depends on its orbit, however.
 Equations (1) for CHZ borders assume a constant amount of insolation. 
They are, therefore, valid for terrestrial planets on circular orbits only. 
The co-existence of a giant and a terrestrial planet in the same 
exoplanetary system necessitates a modification of this concept. 
 Giant planets can alter the orbits of terrestrial planets in non-trivial, time dependent fashion.   This implies that insolation can vary substantially with time. 
For Earth-like planets on variable orbits, the minimum and maximum distances of the planet with respect to its host star determine the maximum and minimum insolation the planet experiences on its  long term orbit. 
Based on long-term radiative equilibrium calculations, the standard climate model (Kasting et al.
1993; Kopparapu et al. 2013, 2014) used to calculate habitable zone limits is not meant to resolve time dependent variations in the amount of insolation.

How quickly an Earth-like planet's climate reacts and adapts to such a time dependent forcing is not completely understood (Bolmont et al. 2016; Popp {\&} Eggl 2017; Way {\&} Georgakarakos 2017).
We, therefore, consider three possible scenarios based on the concept of `climate inertia'.
Zero climate inertia means that the planet's climate reacts quasi instantaneously to changes in insolation. 
This poses strict constraints on the apocenter and pericenter distances a planet can attain in order not to enter a runaway or a freeze-out state.
Previous studies (Williams {\&} Pollard 2002) suggest, however, that a planet's climate can buffer changes in insolation to some degree.
This corresponds to a finite climate inertia.
If the climate inertia tends towards infinity, any variation in insolation can be buffered as long as the insolation average over one orbital period
stays within reasonable (habitable) limits.

\section{Dynamically Informed Habitable Zones}
\subsection{Definitions}
In order to compensate for that deficiency of the climate model, we make use of so-called hereinafter `Dynamically Informed Habitable Zones' (DIHZ) first introduced in Eggl et al. (2012, 2013) and which assume different `climate inertia' of the terrestrial planet.
In this framework, the most conservative estimate regarding the true extent of the HZ, the Permanently Habitable Zone (PHZ), is defined
as the region where a planet with zero climate inertia continuously stays within habitable insolation limits in spite of its possibly elliptic orbit. 
The PHZ is most sensitive to variations in insolation, since it assumes no buffering capabilities
of the atmosphere.  If, for example, the insolation level at the planet's apocenter drops below the CO$_2$ freeze-out level, 
the planet is considered to be uninhabitable, even though it may receive substantially more insolation 
close to its pericenter.
The averaged habitable zone (AHZ), on the other hand, is based on the assumption that the planetary climate can 
buffer any insolation extremum, as long as the insolation average over one orbit remains within habitable limits (Williams {\&} Pollard 2002).
In other words, very high insolation values at the planetary orbit's pericenter and very low insolation values at
apocenter can be disregarded as long as the average amount of light received permits habitability.
Finally, the extended habitable zone (EHZ) is based on limited buffering capabilities of 
the terrestrial planet's climate. Brief excursions to non-habitable insolation levels (one sigma level) are acceptable, 
if the orbital average of the planet's insolation remains within habitable limits. 

\subsection{Calculating Dynamically Informed Habitable Zones}\label{sec:dihz}
Quantifying the concept of climate inertia, dynamically informed habitable zones can deal with elliptic and variable orbits of terrestrial planets.
They can be calculated in the following manner.
The permanently habitable zone (PHZ) can be found by identifying the minimum and maximum distances to the host star along the time dependent orbit.
 For a coplanar hierarchical system that is dynamically stable, changes in orbital semi-major axes are practically negligible and hence the semi-major axes remain constant over long timescales (e.g. Harrington 1968).
Then, the only relevant variable to change with time is $e(t)$. The PHZ borders must not become time dependent themselves, however.
We can solve this issue by searching for $\max_t\; e(t)=e^{max}$,
the estimated maximum eccentricity that the terrestrial planet reaches during its orbital evolution.
The corresponding minimum and maximum distances to the host star then read  $r_{min}=a(1-e^{max})$,
and  $r_{max}=a(1+e^{max})$,
where $a$ is the terrestrial planet semi-major axis.
Consequently, the inner and outer borders of the PHZ can be found by solving the equations 
\begin{equation}
\label{eq:phzio}
S_I=L[a (1-e^{max})]^{-2},\qquad S_O=L[a (1+e^{max})]^{-2},
\end{equation}
with respect to $a$. In a coplanar setup, the solution of equations (\ref{eq:phzio}) yields circles around the host-star defining the edges of the PHZ.
The gravitational influence of the giant planet on the orbit of the terrestrial world depends on the configuration of the system, so that the maximum eccentricity,
although independent of time, is still
a function of other system parameters, such as the terrestrial planet's and the giant planet's semi-major axes ($a,a_g$) and initial eccentricities ($e_{0}, e_{g}$), as well as the masses of the celestial bodies in the system: the stellar mass $m_*$, the giant planet mass $m_g$ and the terrestrial planet mass $m$.
The inner and outer border of the PHZ form an annulus around the host star spanning from $a_I(S_I, a_g,e_{0},e_{g},m,m_g,m_*)$ to $a_O(S_O, a_g,e_{0},e_{g},m,m_g,m_*)$.
In most of the cases, equations (\ref{eq:phzio}) cannot be transformed into an explicit expression for $a$, because $e^{max}$ generally depends on $a$ in a non-trivial manner.
They need to be calculated using numerical techniques for implicit equations. The classical HZ borders can serve as initial starting points for the corresponding algorithm.
Only in the most simple cases explicit expressions can be found (see section A.4 in Appendix A).

In order to find the borders of the averaged habitable zone where we assume the planet has a very high climate inertia, we need to solve
\begin{equation}
\langle S \rangle=S_I, \qquad \langle S \rangle=S_O 
\end{equation}
in terms of $a$, where $S$ is the insolation (in units of Solar constants) that the terrestrial planet receives.
The averaged over time insolation that the terrestrial planet receives is: 
\begin{equation}
\label{eq:sav}
\langle S \rangle=\frac{1}{P_s}\int_0^{P_s}\frac{L}{r^2(t)}dt,
\end{equation}
where we average over $P_s$, the secular period of the terrestrial planet's orbit. Neglecting short period and resonant terms, the planet's orbit evolution returns to its initial state after $P_s$.
Hence, the average over one $P_s$ becomes equivalent to the average over $t\rightarrow\infty$.
For practical reasons, it is more convenient to transform the time average into averages over angular quantities (Arnold 1978). Equation (\ref{eq:sav}) then becomes
\begin{equation}
\label{arnold}
\langle S \rangle\approx\frac{L}{4\pi^2}\int_0^{2\pi}\int_0^{2\pi}\frac{1}{a^2(1-e^2)^{\frac{1}{2}}}df d\phi\approx\frac{L}{4\pi^2}\int_0^{2\pi}\int_0^{2\pi}\frac{1+(1/2)e^2}{a^2}df d\phi=L\frac{1+(1/2)\langle e^2 \rangle}{a^2},
\end{equation}
where we have used Kepler's second law
\begin{equation}
\frac{dt}{r^2(t)}=\frac{df}{na^2(1-e^2)^{1/2}}.\nonumber
\end{equation}
Here, $n$ is the mean motion of the terrestrial planet.  Moreover,
$\langle e^2 \rangle$ is the averaged squared terrestrial planet eccentricity and $\phi=\nu t$,
where $\nu=2\pi/P_s$ is the analytical estimate for the secular frequency of the motion of the terrestrial planet, see section \ref{sec:orbitevolution}. 
The first `approximately equal' sign in equation (\ref{arnold}) occurs due to the fact that the two frequencies $df(t)/dt$ and $\nu$ are not independent. 
However, since the secular timescale is much longer than the orbital period of the terrestrial planet, we have made the assumption that $\nu \ll df(t)/dt$
so that we can consider the system to evolve independently on orbital and secular timescales. 
The second `approximately equal' sign in equation (\ref{arnold}) is used because we carry out a binomial expansion and we only keep the first term of that expansion in $\langle e^2 \rangle$.

The above formula for $\langle S \rangle$ holds for the case of an initially zero planetary eccentricity.
However, it is likely that terrestrial planets form on slightly elliptic orbits due to the interaction between the giant planet and the primordial disc.
More specifically, the initial eccentricity of the terrestrial planet's orbit is dampened towards the forced eccentricity injected by the giant planet (Mardling 2007).  In that case the eccentricity remains constant and the averaged insolation reads
\begin{equation}
\langle S \rangle=\frac{1}{P_s}\int_0^{P_s}\frac{L}{r^2(t)}dt=\frac{L}{2\pi}\int_0^{2\pi}\frac{1}{a^2(1-e^2)^{\frac{1}{2}}}df=\frac{L}{a^2(1-e^2)^{\frac{1}{2}}}.\nonumber
\end{equation}

Finally, the extended habitable zone corresponding to a planet with intermediate climate inertia can be found by solving
\begin{equation}
\langle S \rangle+\sigma=S_I, \qquad \langle S \rangle-\sigma=S_O 
\end{equation}
in terms of $a$.
Hence, in addition to the averaged over time insolation we need to find the insolation variance $\sigma$. 
 
Regarding the variance of the insolation we know that
\begin{equation}
\label{sigma}
\sigma^2=\langle S^2 \rangle-\langle S \rangle^2.\nonumber
\end{equation}
For a terrestrial planet perturbed by a giant planet we have
\begin{eqnarray}
\langle S^2 \rangle&=&\frac{1}{P_s}\int_0^{P_s}\frac{L^2}{r^4(t)}dt\approx\frac{L^2}{4\pi^2}\int_0^{2\pi}\int_0^{2\pi}\frac{(1+e\cos{f})^2}{a^4(1-e^2)^{\frac{5}{2}}}dfd\phi\approx\nonumber\\
&\approx&\frac{L^2}{4\pi^2}\int_0^{2\pi}\int_0^{2\pi}\frac{(1+e\cos{f})^2(1+\frac{5}{2}e^2)}{a^4}dfd\phi=L^2\frac{1+3\langle e^2 \rangle}{a^4},\nonumber
\end{eqnarray}
and consequently
\begin{equation}
\sigma^2=L^2\frac{2\langle e^2\rangle-(1/4)\langle e^2\rangle^2}{a^4}\approx L^2\frac{2 \langle e^2\rangle}{a^4}.\nonumber
\end{equation}

When $e$ is constant, then
\begin{equation}
\langle S^2 \rangle=\frac{1}{P_s}\int_0^{P_s}\frac{L^2}{r^4(t)}dt=\frac{L^2}{2\pi}\int_0^{2\pi}\frac{(1+e\cos{f})^2}{a^4(1-e^2)^{\frac{5}{2}}}df=\frac{L^2}{2a^4}\frac{2+e^2}{(1-e^2)^{\frac{5}{2}}}\nonumber
\end{equation}
and consequently
\begin{equation}
\sigma^2=\frac{L^2}{a^4}\left[\frac{1+(1/2)e^2}{(1-e^2)^{\frac{5}{2}}}-\frac{1}{(1-e^2)}\right].\nonumber
\end{equation}

We should point out here that as a consequence of the way we have defined the PHZ and the EHZ, these two
zones will always be more narrow than the CHZ. From equations (4) and (8) it can be seen that the inner border of 
either the PHZ or the EHZ will be located farther out than the inner edge of the CHZ, while the outer borders
will be inside the outer edge of the CHZ. On the other hand, equations (5) tell us that both borders of the AHZ will be shifted away from the star.  This implies that it is possible in certain cases to have an AHZ that is wider that the CHZ.      

In all the above calculations the stellar radii, luminosities and effective temperatures are assumed constant on the timescale of the orbital
evolution of the terrestrial planet.
Hence, the borders of dynamically informed habitable zones are functions of constant system parameters only and independent of time.
Expressions for $e^{max}$ and $\langle e^2\rangle$ are given in Appendix A.
The equations for PHZ, EHZ and AHZ do not depend on any details of the underlying dynamical model. They work as long as values for $e^{max}$ and $\langle e^2\rangle$ can be provided.
Similarly, limits for the habitable insolation values $S_I$ and $S_O$ can be chosen differently from those used (Kopparapu et al. 2014) in this work,  if so desired.  This makes the here presented method extremely versatile.
Figure 1 contains a graphical representation of DIHZ for the HD150706 system.

\begin{figure}
\begin{center}
\includegraphics[width=100mm]{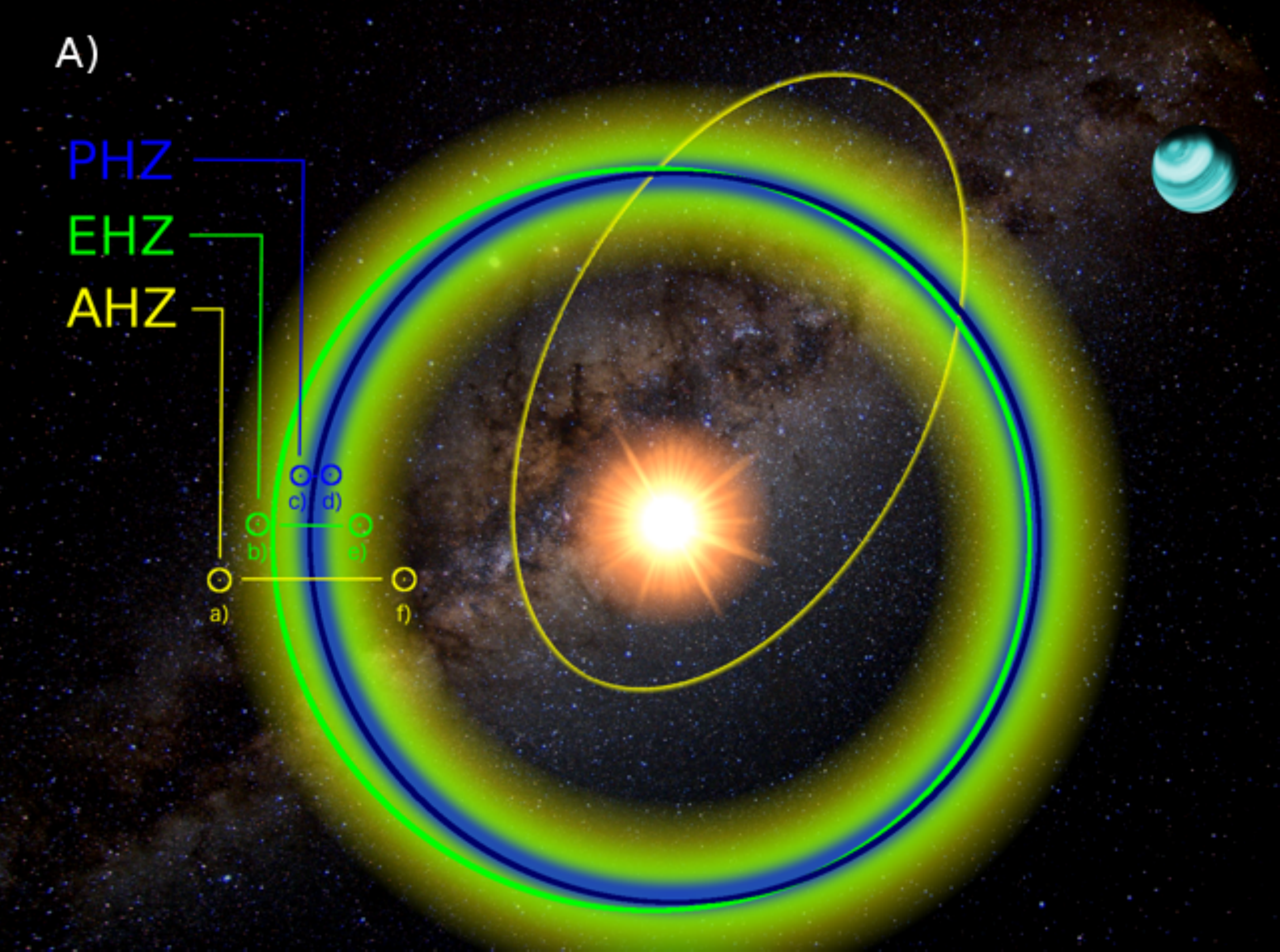}\\
\includegraphics[width=80mm]{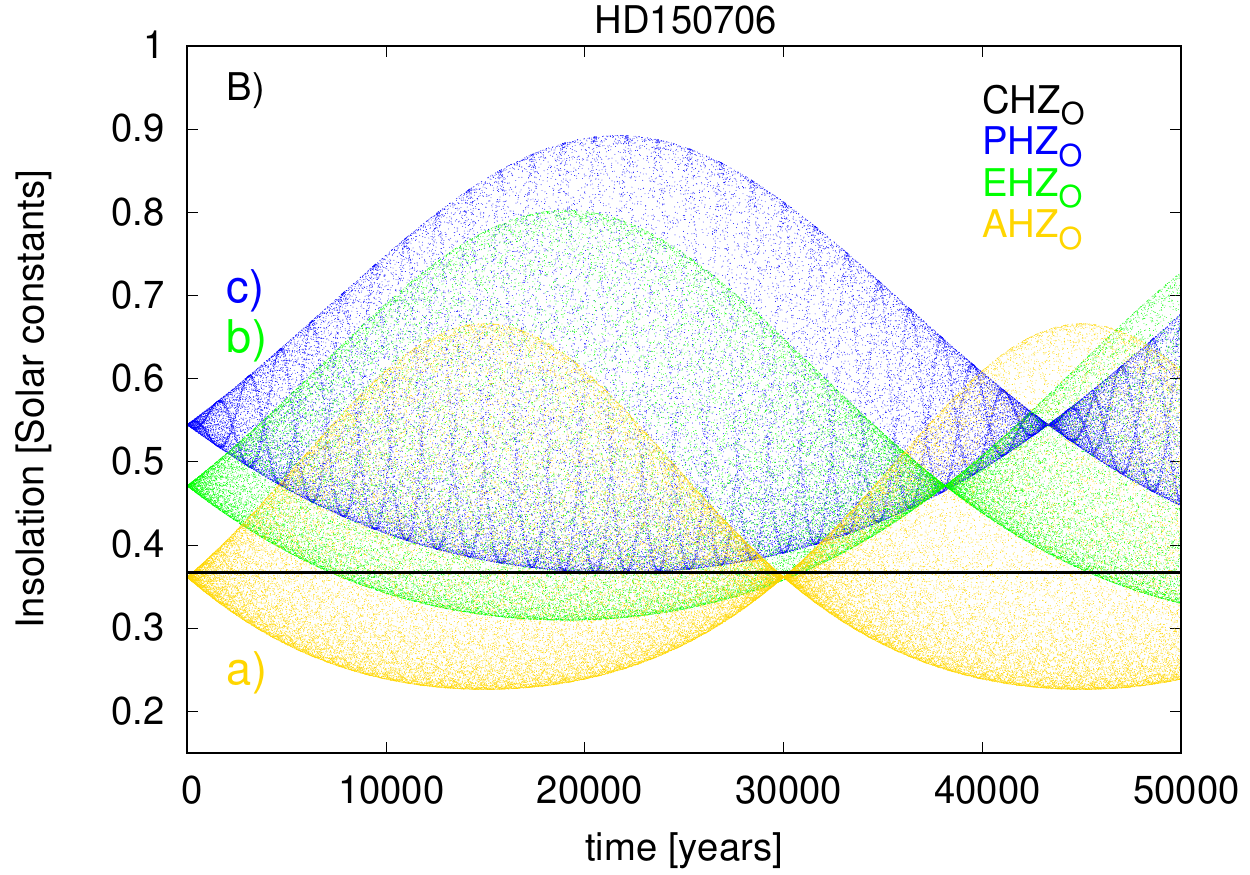}
\includegraphics[width=80mm]{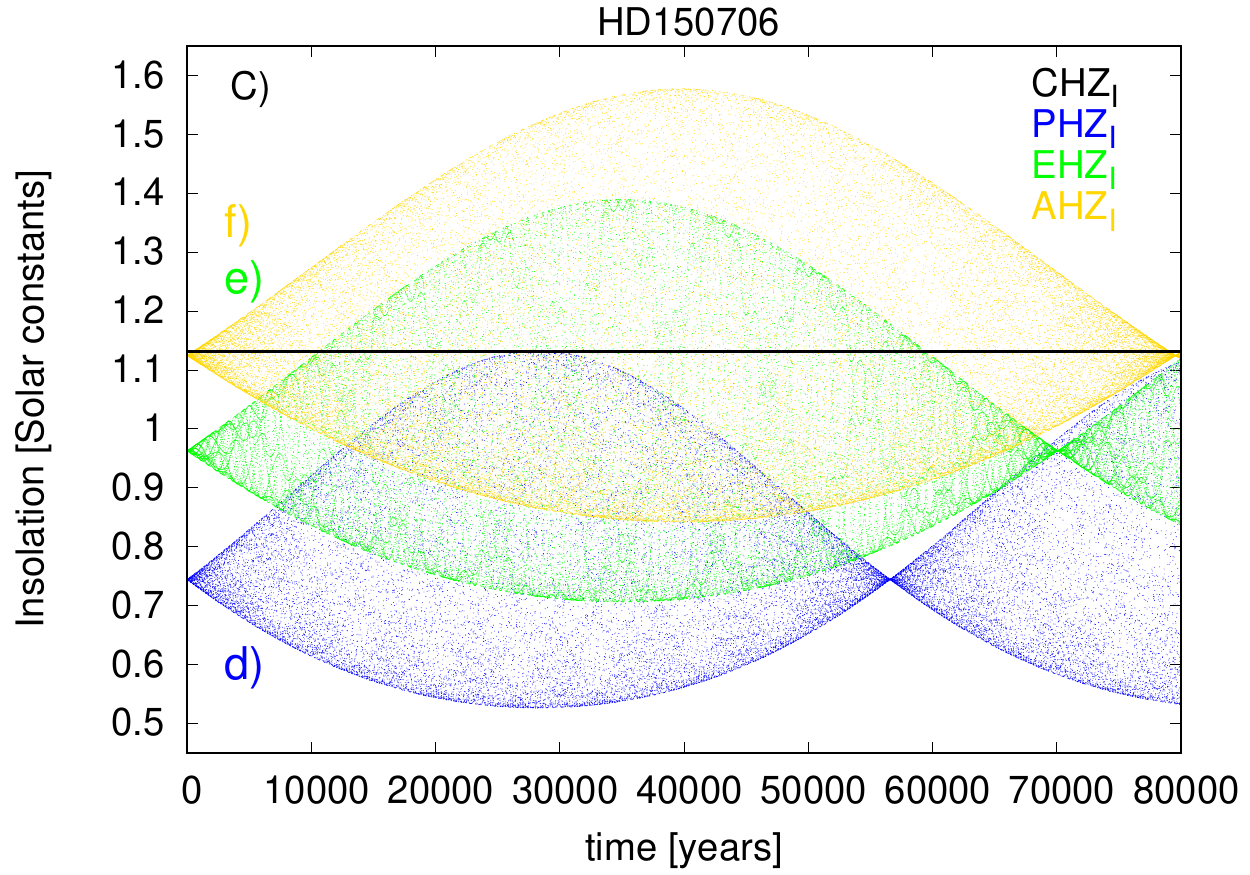}\\
\caption{DIHZs for the HD150706 system. Panel A shows the three dynamically informed habitable zones, PHZ (blue), EHZ (green) and AHZ (yellow) in the HD150706 system (not to scale).  Panel B demonstrates the insolation evolution of a 
fictitious additional planet orbiting at the outer border of the dynamically informed habitable zones. Panel C is the same as panel B but for the inner border. The black line represents the insolation that the fictitious planet would receive at the borders of the CHZ.}
\end{center}
\end{figure}

\subsection{Dynamically informed vs classical habitable zones}\label{sec:dihzvschz}
Do the changes in the orbit of a terrestrial planet induced by a gas giant merit the additional effort of calculating dynamically informed habitable zones?
In order to answer this question, we can determine the percentage relative displacement of the PHZ borders with respect to the classical HZ borders:
\begin{equation}
\label{rdi}
D_{I,O}=100\frac{|r_{I, O}-PHZ_{I, O}|}{r_{I, O}}\nonumber
\end{equation}
Using equations (1) and (4), we find
\begin{equation}
\label{rdo}
D_I=100\frac{e^{max}}{1-e^{max}}, \qquad D_O=100\frac{e^{max}}{1+e^{max}}\nonumber.
\end{equation}
We choose the PHZ for the comparison as it provides the most conservative estimates of all the types of habitable zones.
Planets in the PHZ are the most likely to be habitable. 
Figure 2 is a graphical representation of the percentage relative displacement for both inner and outer habitable zone borders assuming a giant planet at various distances in S-type  (the giant planet's pericenter distance is larger than the distance of the outer border of the classical habitable zone) and P-type configuration (the apocenter distance of the giant planet is smaller than the distance of the inner border of the classical habitable zone). We use the lowest order secular model and we also assume that the stellar mass is much larger than the planetary masses; then the expression for $e^{max}$ is independent of the masses of the system (see section A.2). 
The displacement increases as $a_g/r_{I, O}$ gets smaller for an S-type system or as $r_{I, O}/a_g$ decreases when we deal with a P-type system. Higher eccentricities of the giant planet's orbit also shrink the PHZ with respect to the classical HZ.
In fact, even at a distance of 30 a.u. a Jupiter-sized planet with  $e_g>0.8$ can displace the HZ borders around a sunlike star by up to  50\%.
It is worth noting that for a given distance of the giant planet to the Earth-analog, terrestrial planets close to the inner edge of the HZ are more severely affected by the gravitational influence of the giant planet than Earth-like worlds close to the outer edge of the HZ.
This is due to the fact that the magnitude of the insolation gradient $|\nabla S(r)|\propto r^{-3}$ is steeper closer to the host star causing larger changes in insolation for the same orbital eccentricity. 
Figure 2 becomes less accurate for smaller semi-major axis ratios (e.g. lower than 4), as the plots were constructed using only the lowest order secular model.  
For semi-major axis ratios lower than 4 the higher order model mentioned in section A.2 would be more appropriate. Similarly, resonances start to become important as well in this region so that a more complete dynamical model should be used.   

\begin{figure}[H]
\renewcommand{\baselinestretch}{1.}
\includegraphics[width=80mm,height=50mm]{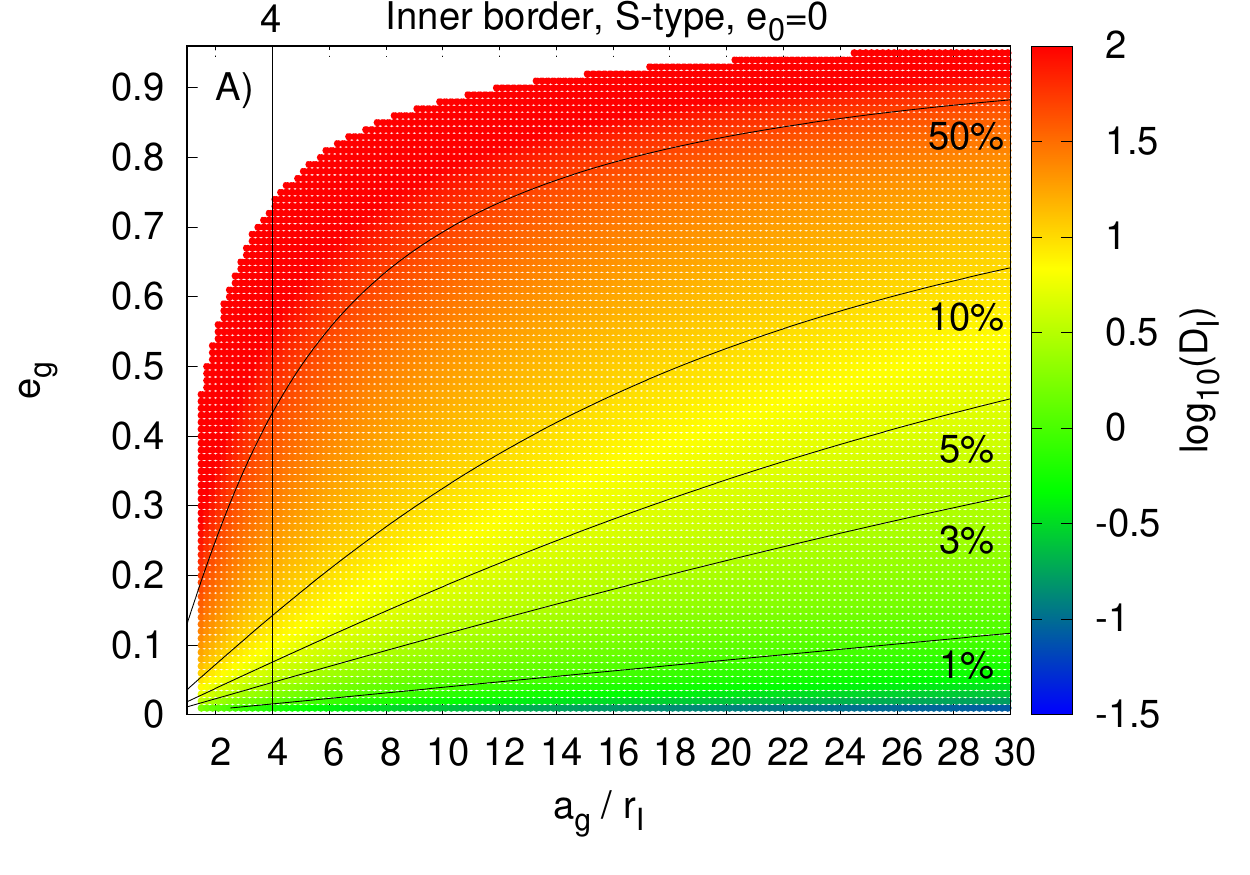}
\includegraphics[width=80mm,height=50mm]{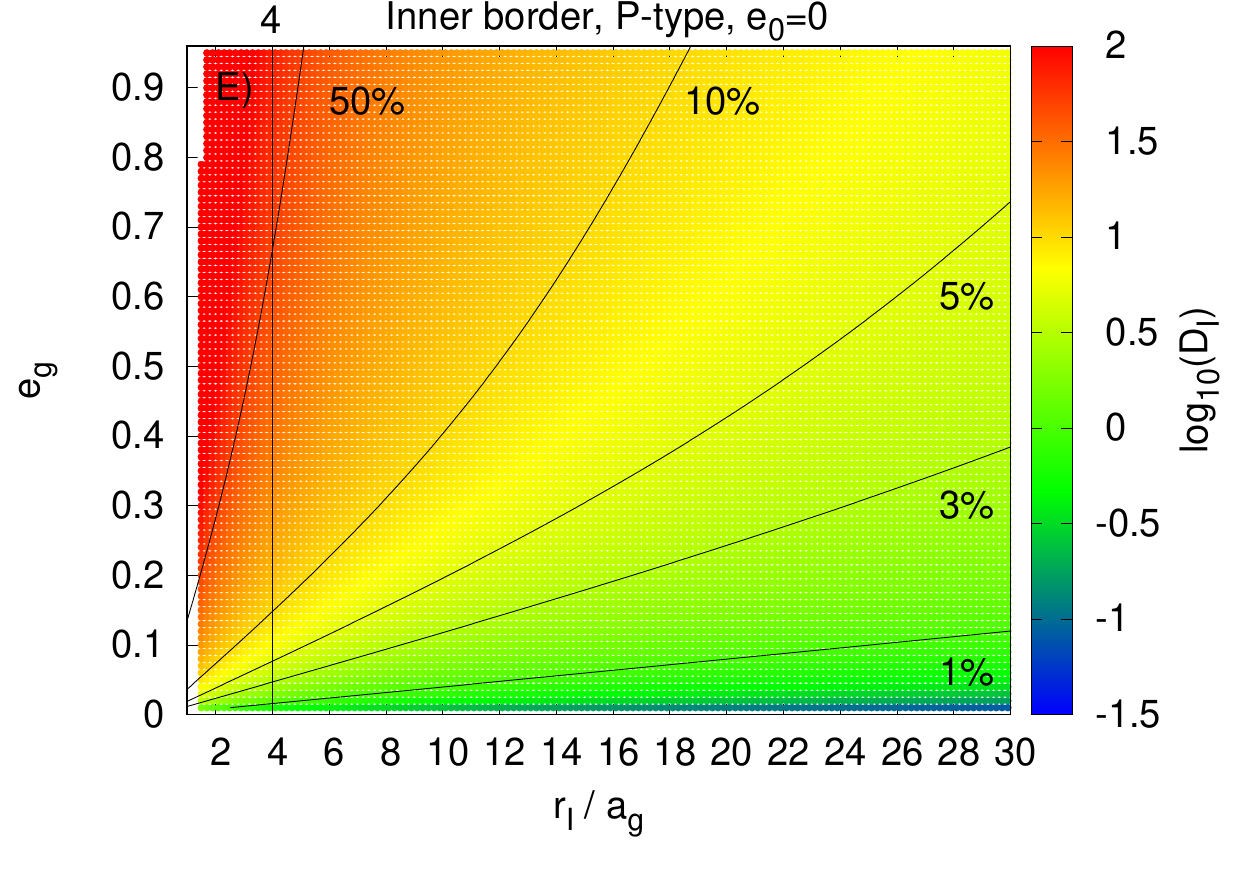}
\includegraphics[width=80mm,height=50mm]{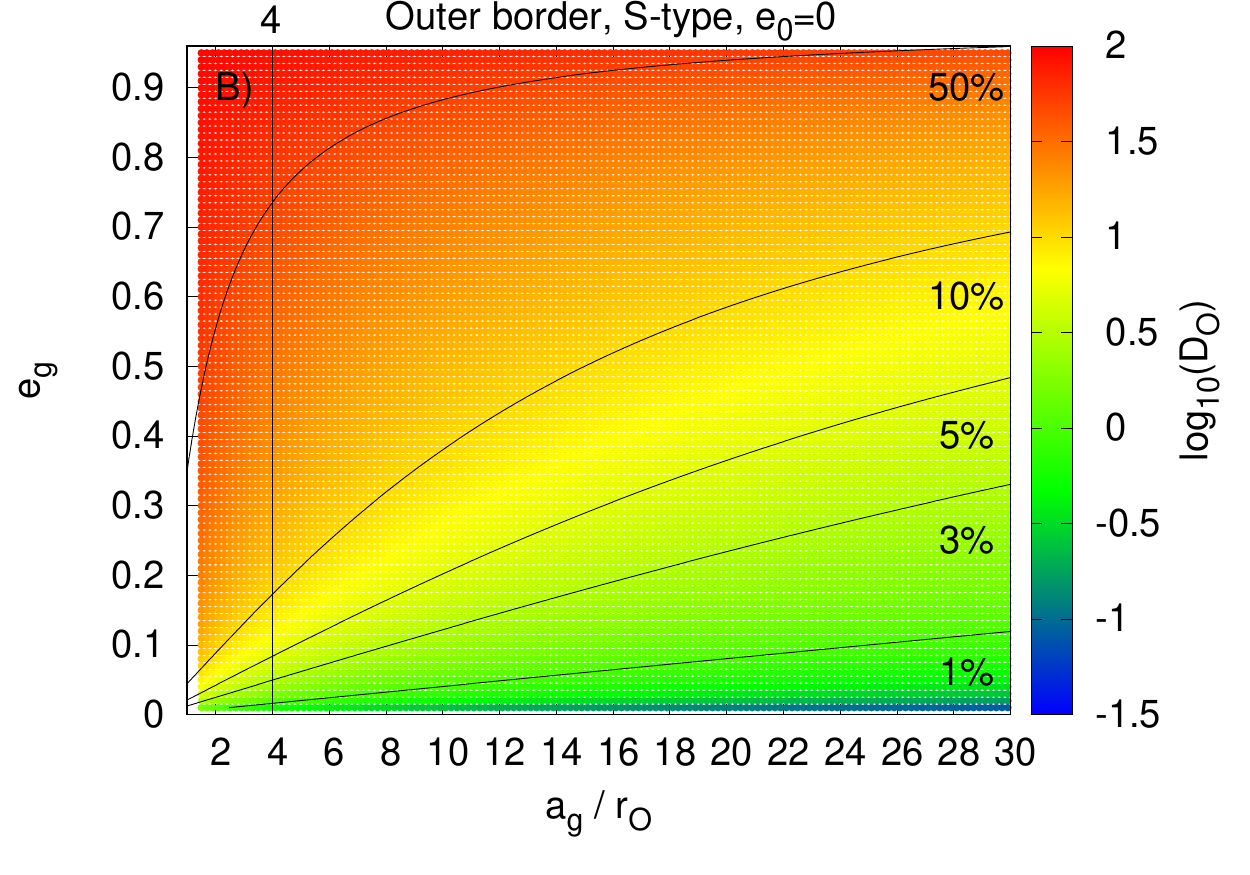}
\includegraphics[width=80mm,height=50mm]{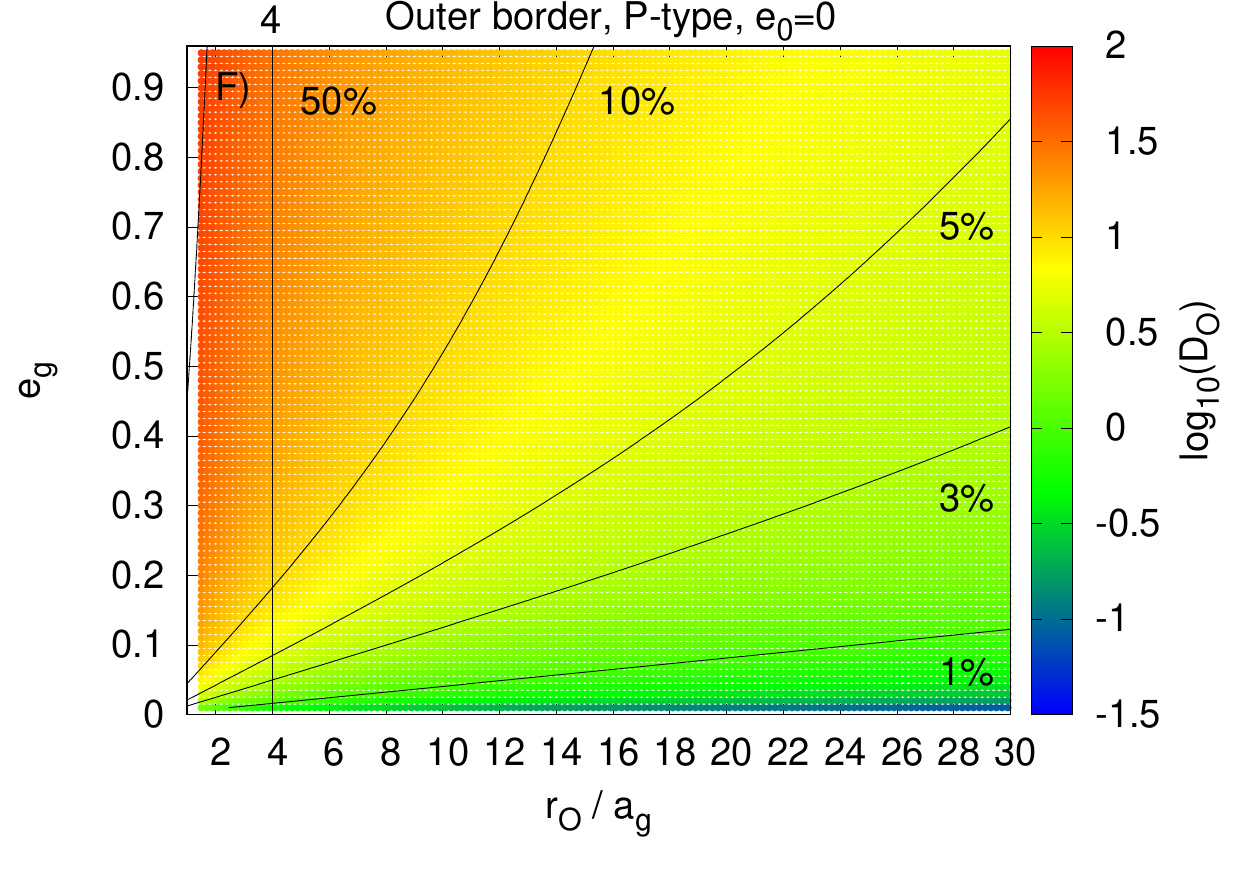}
\includegraphics[width=80mm,height=50mm]{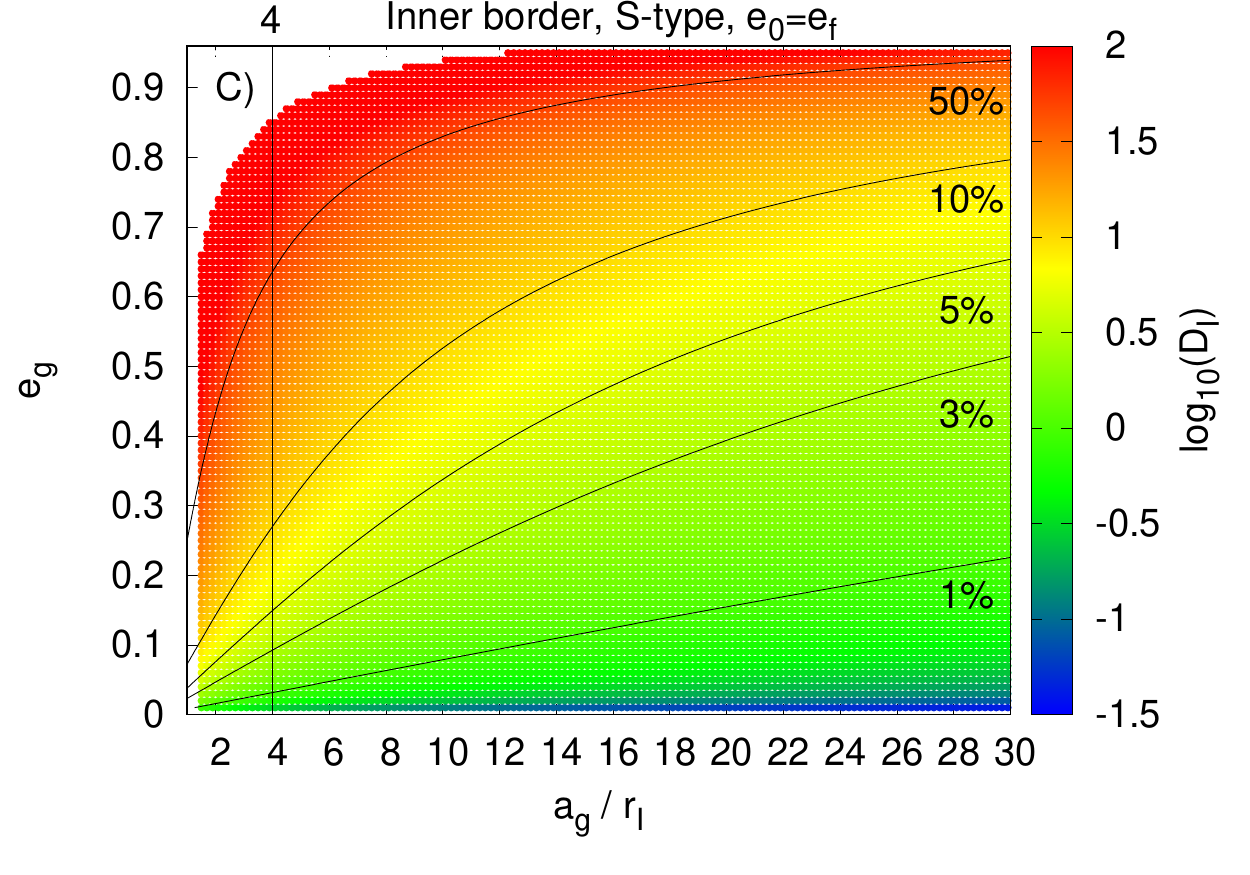}
\includegraphics[width=80mm,height=50mm]{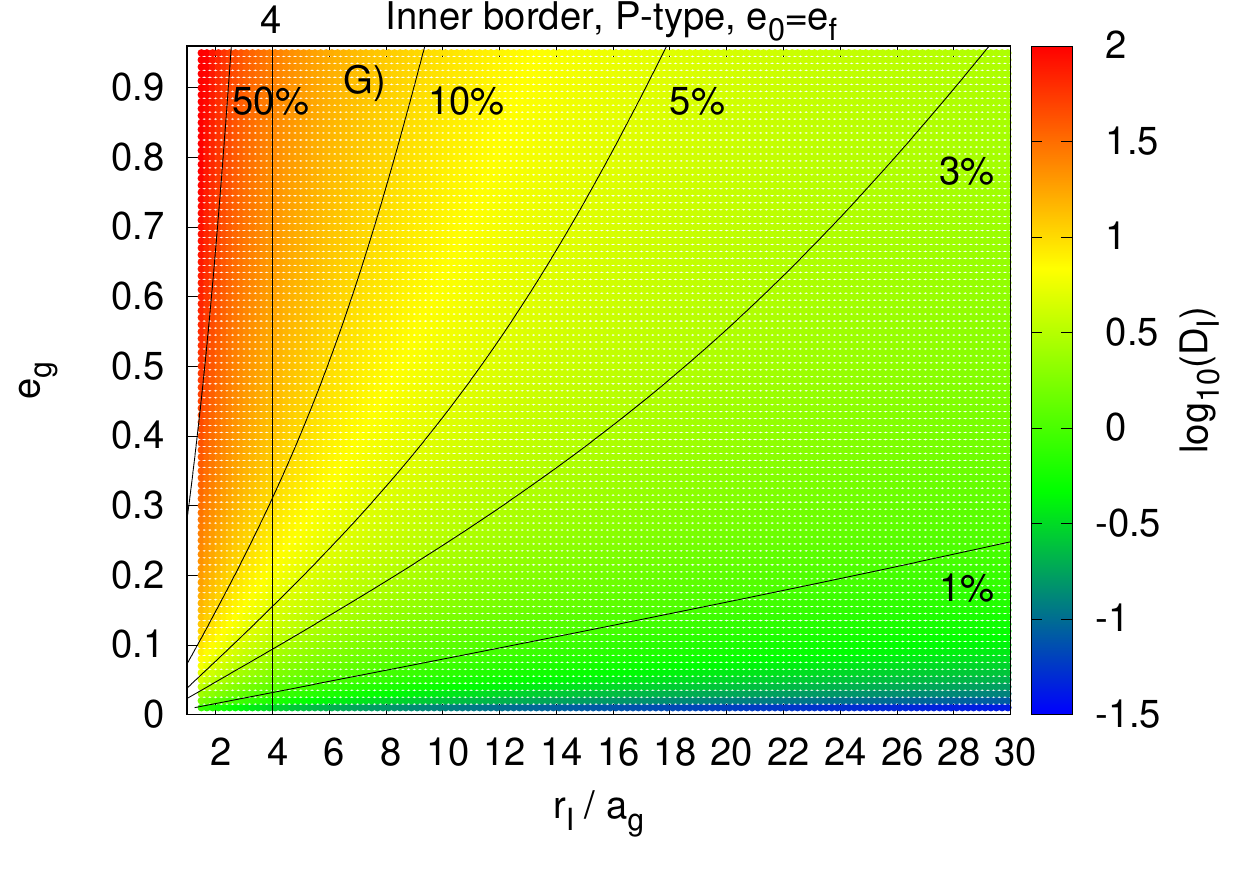}
\includegraphics[width=80mm,height=50mm]{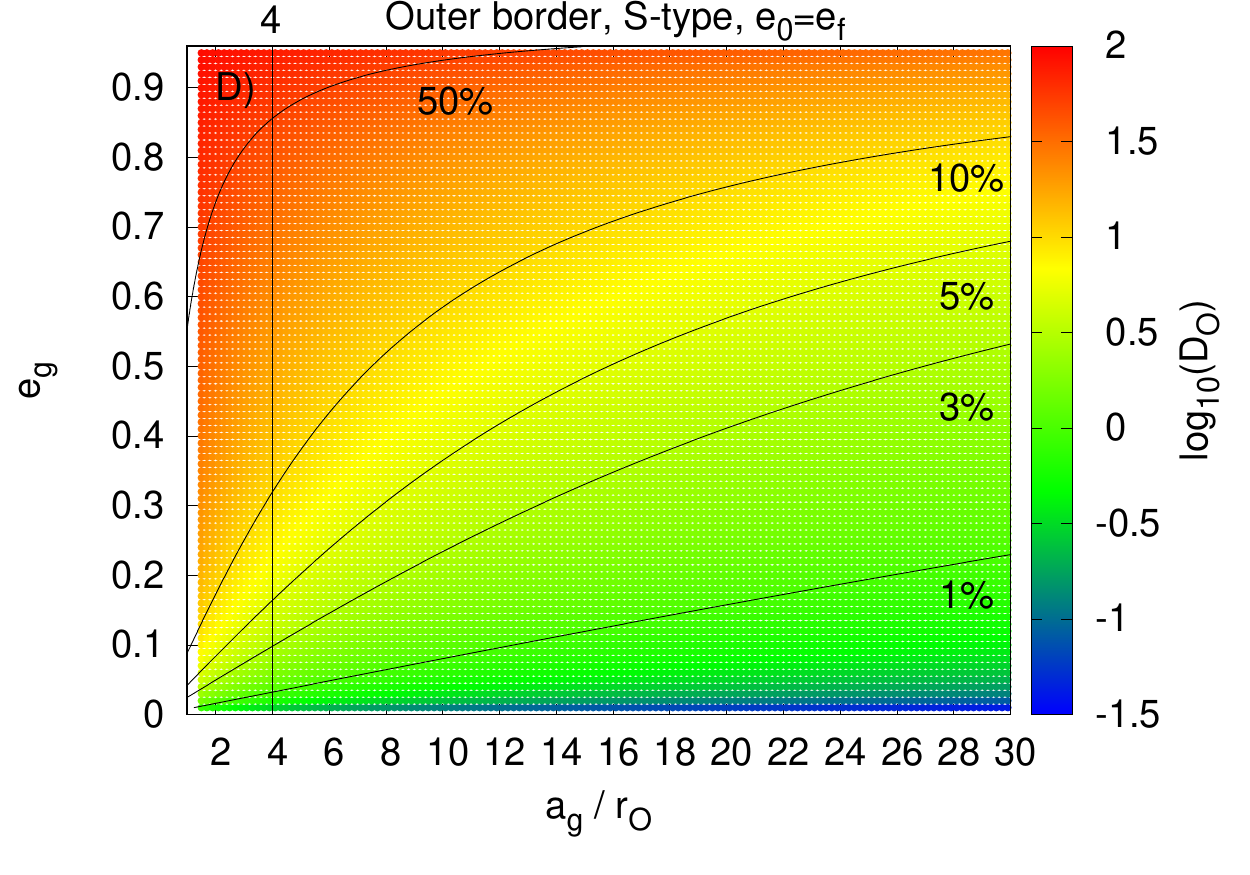}
\hspace{1.6cm}\includegraphics[width=80mm,height=50mm]{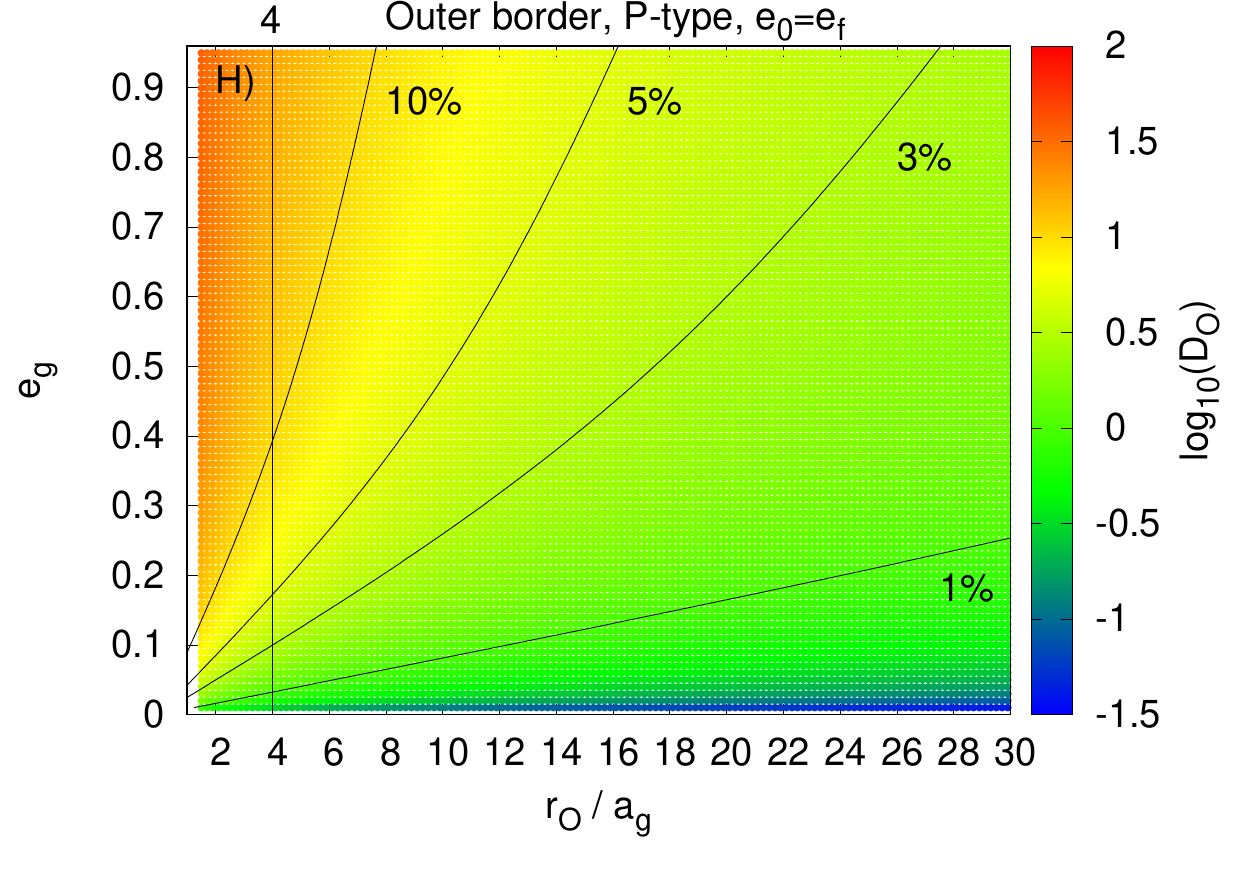}
\caption{Percentage relative displacement of the boundaries of the CHZ. The displacement, which is denoted by color (in logarithmic units), is with respect to the PHZ boundaries. Panels A, B, C and D refer to S-type systems, while panels E, F, G and H refer to P-type systems.  
The five black curves correspond from bottom to top to $D=1\%$, $D=3\%$, $D=5\%$, $D=10\%$ and $D=50\%.$\label{fig:disp}}
\end{figure}

\section{Application to known exoplanet systems}\label{sec:application}
 
The DIHZ are used to assess which of the currently known exoplanet systems hosting a main sequence star and a giant planet offer the highest chances of finding additional, potentially habitable worlds.  Out of the known exoplanet population, we have selected 147 systems consisting of a star on the main sequence and a single giant planet with well determined orbital parameters.
We choose giant planets with masses in the range of $0.1 M_J-13 M_J$, $M_J$ being the mass of Jupiter and stars with an effective temperature $2600\leq T_{eff}\leq 7200K$ and $R_* \leq 1.3 R_{\odot}$.  
Our sample does not contain systems where the giant planet crosses the borders of the classical habitable zone as 
they could harbor habitable moons or Trojan planets, both of which require special treatment.
Assuming the presence of an additional, fictitious Earth-like planet, we investigate the effect of the actual giant planet on the former if it were orbiting its host
star in the vicinity of the system's classical habitable zone.

In our sample, we have 131 P-type configurations and 16 S-type systems.
Classifying the systems based on the orbital period of the giant planet, we have 123 hot Jupiters ($T_{g}<$ 10 days, 
$T_{g}$ being the orbital period of the giant planet), 9 warm Jupiters (10 days $<T_{g}<$ 400) and 16 cold Jupiters ($T_{g}>400$). The selected systems together with their key parameters and corresponding uncertainties can be found in the Appendix B.
In order to assess the influence a giant planet exerts on a terrestrial planet and its habitability 
we calculate the extent of the classical habitable zone and compare it to the
extent of dynamically informed habitable zones.  This comparison allows us to quantify the effect of the giant planet on the habitability of additional terrestrial planets yet to be discovered in those systems.  For the borders of the CHZ, we have used the runaway greenhouse and the maximum greenhouse insolation limits.
The percentage shrinkage ($\Sigma$) 
\begin{equation}
\Sigma(\text{DIHZ})=100\left(1-\frac{\text{DIHZ}_O-\text{DIHZ}_I}{r_O-r_I}\right),
\end{equation}
quantifies that notion, where DIHZ stands for the respective dynamically informed habitable zone (PHZ, EHZ or AHZ). As before the subscripts $I$ and $O$ 
denote the inner and outer border, respectively.
If atmospheric buffering plays a large role in determining whether a system is habitable or not, 
values for $\Sigma$(PHZ) and $\Sigma$(AHZ) will differ substantially. Negative shrinkage values mean the respective dynamically informed habitable zone is 
larger than the classical habitable zone.

Figure 3 is a graphical representation of the habitable zone shrinkage as given in equation (9) for the systems in our sample, where we have assumed that the terrestrial planet occupies the least excited dynamical state after the formation phase, 
i.e. its orbit's eccentricity corresponds to the forced eccentricity. 
\begin{figure}
\begin{center}
\includegraphics[width=89mm]{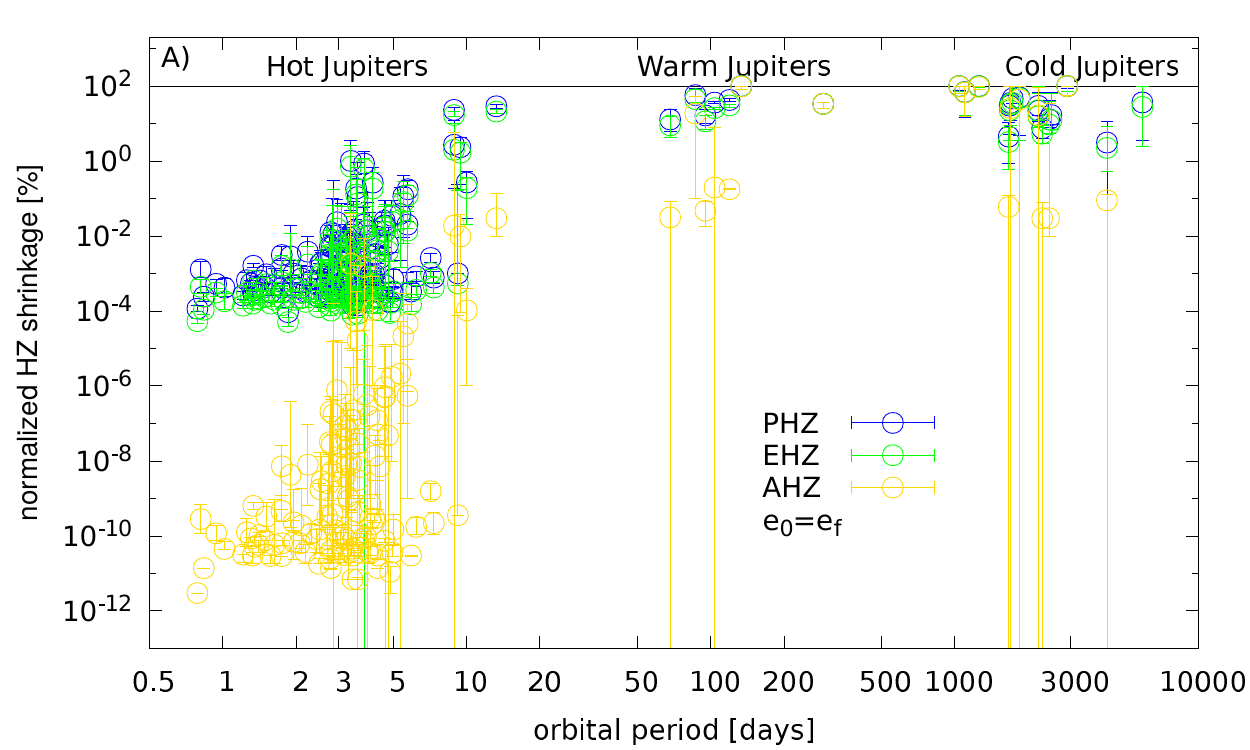}
\includegraphics[width=89mm]{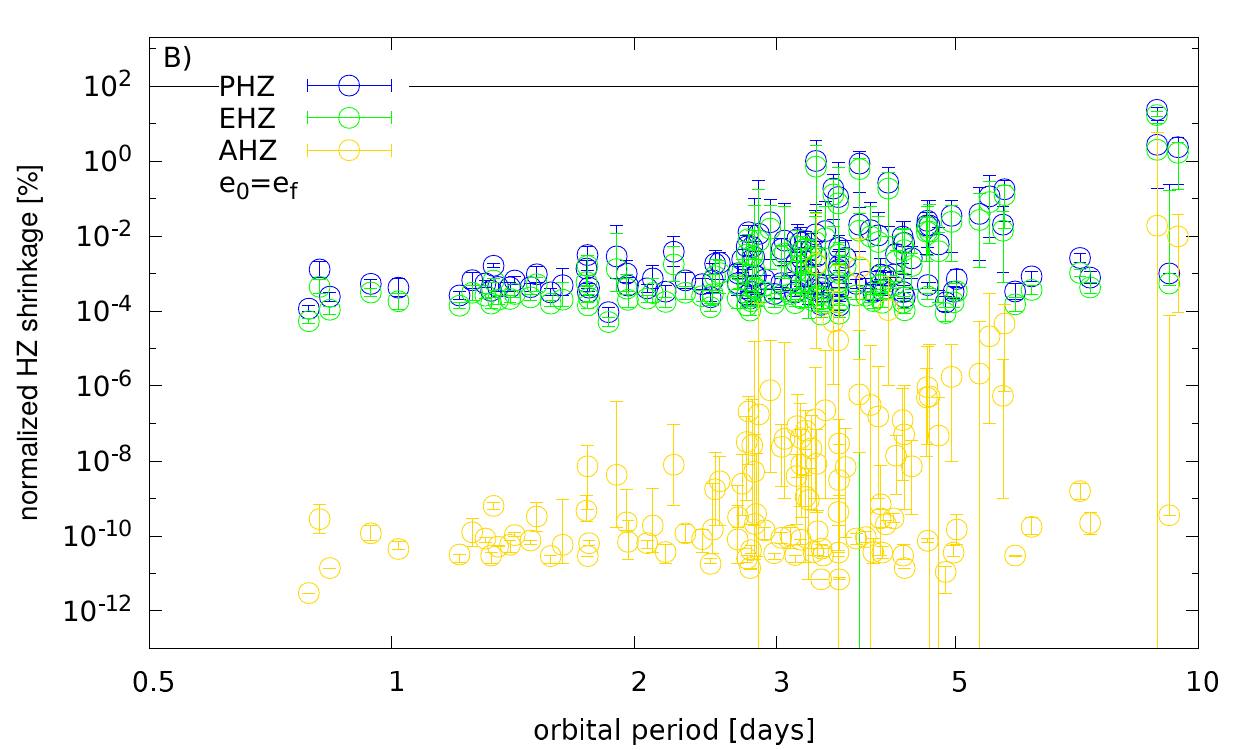}
\includegraphics[width=89mm]{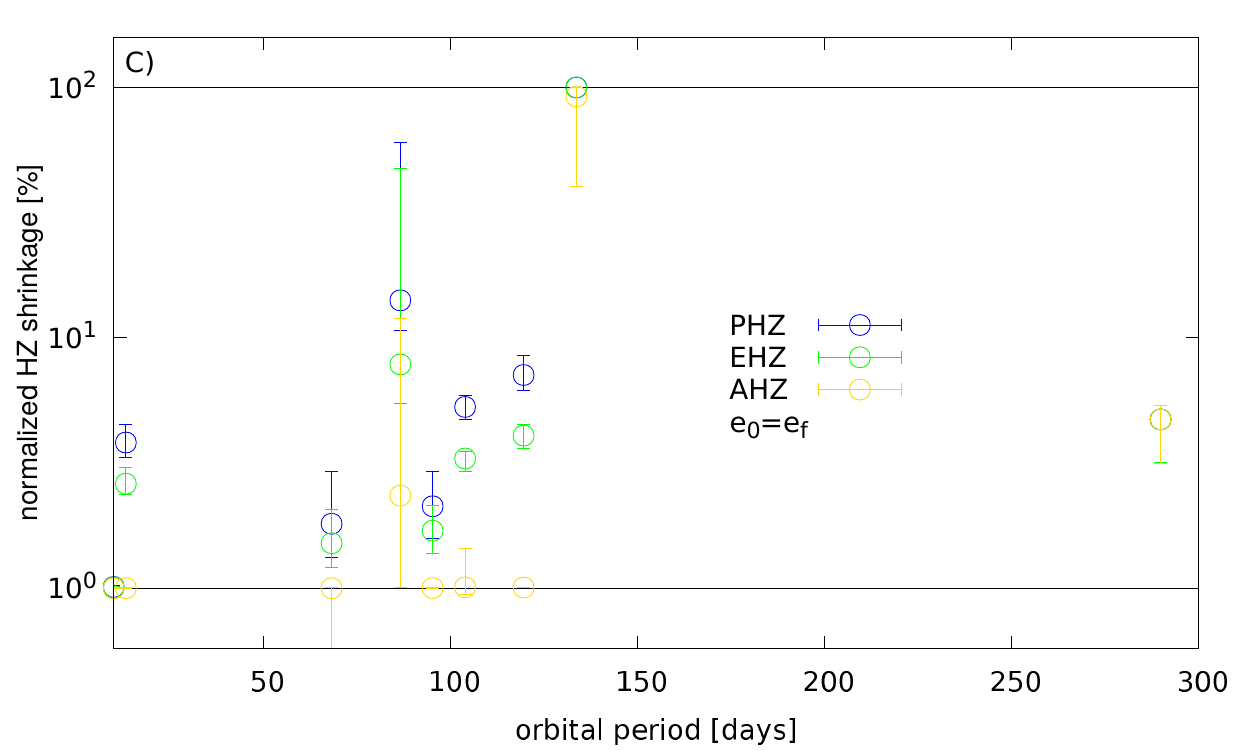}
\includegraphics[width=89mm]{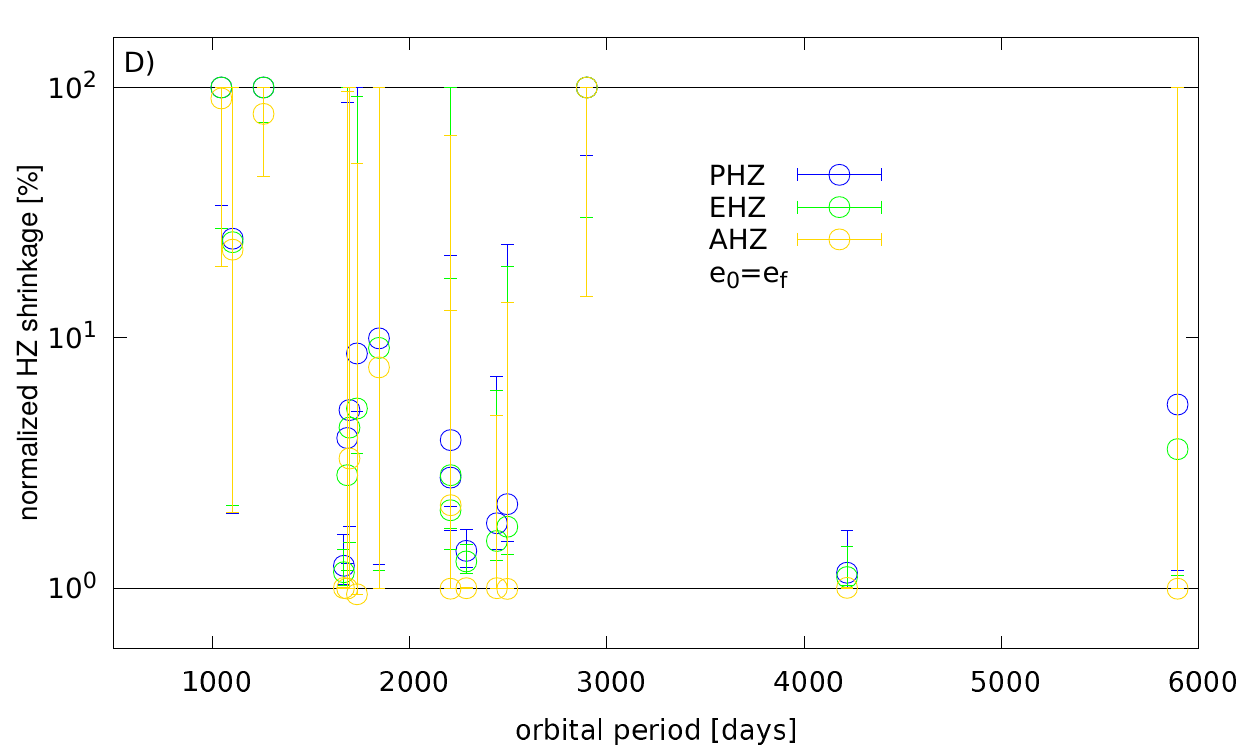}
\caption{Habitable zone shrinkage against the giant planet's orbital period. Panel A presents the results for all our systems, while panels B, C and D show the shrinkages for hot, warm and cold Jupiter systems respectively. All shrinkage values are given with their corresponding uncertainties. Also $e_{0}=e_f$. \label{fig:shr1}}
\end{center}
\end{figure}
Panel B in figure 3 shows that hot Jupiters orbiting close to their host star have very little 
influence on a system's capability to host habitable worlds. In most hot Jupiter systems $\Sigma$(DIHZ)$\ll1\%$.
While terrestrial worlds seem to be sensitive to atmospheric buffering capabilities in such environments - there are several orders of magnitude differences
between AHZ and PHZ shrinkage values - both PHZ and AHZ shrinkage are insignificant in absolute terms, however. 
This is to be expected as most of the giant planets have zero or nearly zero eccentricities due to tidal interactions with the host star, resulting in 
very small perturbations on the orbit of the terrestrial world.
Such systems, therefore, appear to be good candidates for follow up observations attempting to find additional habitable worlds. 
Moreover, the classical habitable zone is an excellent approximation to the actual habitable region around the host star
in systems with one hot Jupiter on a short period orbit ($P_g<3$ days). This assumes, of course, that no other planets populate the system.  In systems with hot Jupiters on orbits with periods between 3 and 5 days habitable zones are somewhat smaller with a larger scatter compared 
to systems where giant planets have shorter orbital periods.
This is a direct consequence of the larger variety of orbital eccentricities of those giant planets.
Increased orbital eccentricity in giant planets close to their host star 
may be attributed to an unseen companion or less efficient tidal damping due to the larger distance from the host star. 
Higher giant planet eccentricities result in higher maximum and averaged square eccentricity for fictitious terrestrial planets in the system
which, in turn, cause smaller DIHZs. 

Those systems in our sample that contain warm or cold Jupiters exhibit the greatest shrinkage of the actual 
region where an Earth-like planet can harbor liquid water near its surface.  In fact, 
in certain cases the habitable zones are completely eliminated (shrinkage of $100\%$) due to either 
a substantial growth in the orbital eccentricity of the terrestrial planet violating insolation constraints for habitability, 
or complete dynamical instability of the CHZ.
Warm and cold Jupiter systems are also more sensitive to climate inertia as demonstrated by the greater differences between AHZ and PHZ.   
Although such systems do not appear to be the most attractive targets in the search for habitable worlds, further interpretation is required here. 
There is a clear observational bias in our sample towards giant planets with short orbital periods and high eccentricities. 
Cold Jupiters on circular orbits will affect habitable zones significantly less than similar planets on eccentric orbits. As the former are more 
difficult to discover, our sample becomes very sparse in that domain. We, therefore, have to resort to theory-derived arguments rather than
observational data to assess the consequences for habitable zones in systems where the giant planet orbits its host star at a greater distance.
As seen in subsection 4.3, we find that cold Jupiters well beyond the habitable zone (e.g. $a_g/HZ_I > 4$) on orbits with eccentricities lower
than $e_g<0.3$ do not substantially influence potentially habitable planets.  
In our Solar System, for instance, Jupiter is located at $5.2$ a.u. with an eccentricity of $e_g\approx0.05$. Assuming an initially circular orbit for the Earth,  
this translates into a $\Sigma$ (PHZ) of $12\%$.
If Jupiter had an $e_g\approx0.1$ eccentricity instead, the border displacements would double leading to a PHZ shrinkage of around $22\%$.

A glance at two systems from our sample, HD13931 and HD150706,  confirms that higher eccentricities of giant planets can cause a substantial reduction of a system's habitable zone, even if the giant planets are relatively far out.
HD13931 hosts a giant planet roughly three times as massive as Jupiter on an almost circular orbit ($e_g\approx 0.02$ ) with a semi-major axis of  $5.15$ a.u. and an orbital period of 4218 days. 
The fact that HD13931b orbits a sunlike star makes this systems very similar to our Solar System. 
Panel D of figure \ref{fig:shr1} shows that, unlike most of the other cold Jupiter systems in our sample, HD13931 suffers relatively little HZ shrinkage, at most $\Sigma$ (PHZ) $\approx 23\%$.
The discrepancy between PHZ and AHZ is low as well, indicating that atmospheric
buffering capabilities of additional terrestrial worlds are not crucial in this system.
Thus, HD13931 is a good candidate for follow up observations. 
HD150706b on the other hand, orbits its sunlike host-star at a nominal distance of 6.7 a.u. with an orbital eccentricity of 0.38. The planet has an expected mass
around 9 $M_J$ and an orbital period of a little more than 16 years. 
The nominal permanently habitable zone shrinkage $\Sigma$(PHZ) in this system lies between 37\% ($e_0=e_f$) and 72\% ($e_0=0$) depending on the dynamical state of the terrestrial planet. However, uncertainties in the system parameters allow the PHZ to vanish completely, i.e. $\Sigma$ (PHZ) =100\%. 
Even if the terrestrial planet's atmosphere has a limited capability to buffer changes in insolation
only roughly two thirds of the classical habitable zone could support a planet with liquid water on its surface, since the nominal $\Sigma$(EHZ)$\approx$28\% for ($e_0=e_f$) 
and 38\% for ($e_0=0$).  
If the Earth-like planet has a high climate inertia instead, the HD150706 system offers a slightly larger AHZ compared to classical HZ estimates
since $\Sigma$(AHZ)$\approx$-0.1\% for ($e_0=e_f$) and -1.5\% for ($e_0=0$),
albeit at a greater distance from the host star. 
HD150706 is not the only system experiencing a slight enlargement of the AHZ compared to a shrinkage in the other dynamically informed habitable zones.
About 6\% of the systems in our sample have a larger averaged habitable zone than suggested by classical estimates under the condition that the atmosphere of an 
additional terrestrial world is capable of buffering variations in insolation.

All in all, from the gravitational interactions between the terrestrial planet and the giant planet point of view, systems with giant planets on almost circular orbits seem 
to be the best places to look for habitable worlds independent of the climate inertia of the latter. 
Tidal interactions between the giant planet and the host star
naturally create these conditions for close-in hot Jupiters. The fact that those systems have largely unperturbed habitable 
zones may be counterbalanced by recent results suggesting that hot Jupiters could have fewer sub-Jovian companions compared to giant planets farther from the host star (Steffen et al. 2012; Huang et al. 2016). Table 1 contains those systems from our sample that are suitable for follow-up observations. Most of the selected systems have 
habitable zone shrinkages of under $1\%$, while few of them have a maximum shrinkage of the PHZ of the order of $10\%$.
More observational data is needed, however, to correctly estimate occurrence rates for Earth-like companions in systems
with a gas giant and assess their effect on habitability.

\begin{table}[H]
\caption{Candidates for follow-up observations.} 

{\begin{tabular}{c c c}\hline
& Name  &  \\
\hline\\
&BD-114672, CoRoT-12, CoRoT-13, CoRoT-14, CoRoT-16, CoRoT-18, CoRoT-2,CoRoT-25, CoRoT-27, &\\ 
&CoRoT-29, CoRoT-4, CoRoT-5, CoRoT-6, CoRoT-8,HAT-P-12, HAT-P-18, HAT-P-19, HAT-P-21, &\\
&HAT-P-22, HAT-P-23, HAT-P-25, HAT-P-27, HAT-P-28, HAT-P-29, HAT-P-3, HAT-P-36, HAT-P-37, &\\
&HAT-P-38, HAT-P-43, HAT-P-5, HAT-P-51, HAT-P-52, HAT-P-53, HAT-P-54, HAT-P-55, HATS-1, &\\
&HATS-10, HATS-13, HATS-14, HATS-15, HATS-16, HATS-18, HATS-2, HATS-25, HATS-28, HATS-29, &\\
&HATS-30, HATS-32, HATS-33, HATS-34, HATS-4, HATS-5, HATS-8, HD13931, HD63454, K2-29, K2-30,&\\
&K2-31, Kepler-15, Kepler-17, Kepler-41, Kepler-423, Kepler-425, Kepler-426, Kepler-428, & \\
&Kepler-63, Kepler-74, Kepler-77, Qatar-1, Qatar-2, TrES-3, WASP-101, WASP-104, WASP-117, &\\
&WASP-119, WASP-123, WASP-124, WASP-126, WASP-129, WASP-132, WASP-135, WASP-139, WASP-140, &\\
&WASP-157, WASP-16, WASP-18, WASP-19, WASP-21, WASP-23, WASP-25, WASP-26, WASP-28, WASP-29, & \\
&WASP-31, WASP-32, WASP-34, WASP-35, WASP-37, WASP-39, WASP-43, WASP-44, WASP-49, WASP-5, &\\
&WASP-50, WASP-52, WASP-56, WASP-58, WASP-6, WASP-60, WASP-62, WASP-64, WASP-65, WASP-67, &\\
&WASP-69, WASP-75, WASP-80, WASP-83, WASP-89, WASP-95, WASP-96, WASP-97, WTS-1, WTS-2, XO-5 & \\
&-----------------------------------&\\
&Comments&\\
& BD-114672 and HD13931 are the only S-type systems&\\
&BD-114672, CoRoT-6, HAT-P-54, HD13931, Kepler-63 and WASP-89 have a maximum PHZ shrinkage of $\sim 10\%$&\\
&The rest of the systems have a maximum PHZ shrinkage less than $1\%$&\\
\hline
\end{tabular}}
\end{table}

\subsection{The role of the initial orbit eccentricity of the terrestrial planet}
The results presented in figure 3 were calculated assuming dynamically relaxed systems, where $e_{0}=e_f$. However, a terrestrial planet is not guaranteed to end up in such a state after the planet formation phase is over. 
In fact, it will most likely experience larger orbit perturbations than those assumed when plotting figure 3.
In order to see how robust the presented results are against variation in the initial orbital eccentricity of potentially habitable planets
we have examined habitable zone shrinkages for $e_{0}=0$ as well.  
The results were qualitatively similar to the case of $e_{0}=e_f$, although shrinkage values were higher, as was to be expected.  
Especially the extent of the PHZ decreases in dynamically excited systems, since $e^{max}$ can increase by $50\%$ compared to
the most relaxed dynamical states (see equations A1 and A2 in section A2).
Figure 4 shows $\Sigma$(PHZ) and $\Sigma$(EHZ) for dynamically relaxed ($e_{0}=e_f$), and
excited ($e_{0}=0$) systems. As expected, the PHZs and EHZs are smaller in excited systems. 
Note that the scale of the axis indicating the shrinkage is logarithmic.
One can also see that EHZ and PHZ show very similar behavior, indicating that only extremely high climate buffering capabilities (AHZ)
protect a terrestrial planet against variations in insolation that are a consequence of the orbit perturbations induced by the giant planet.

\begin{figure}[H]
\begin{center}
\includegraphics[width=105mm]{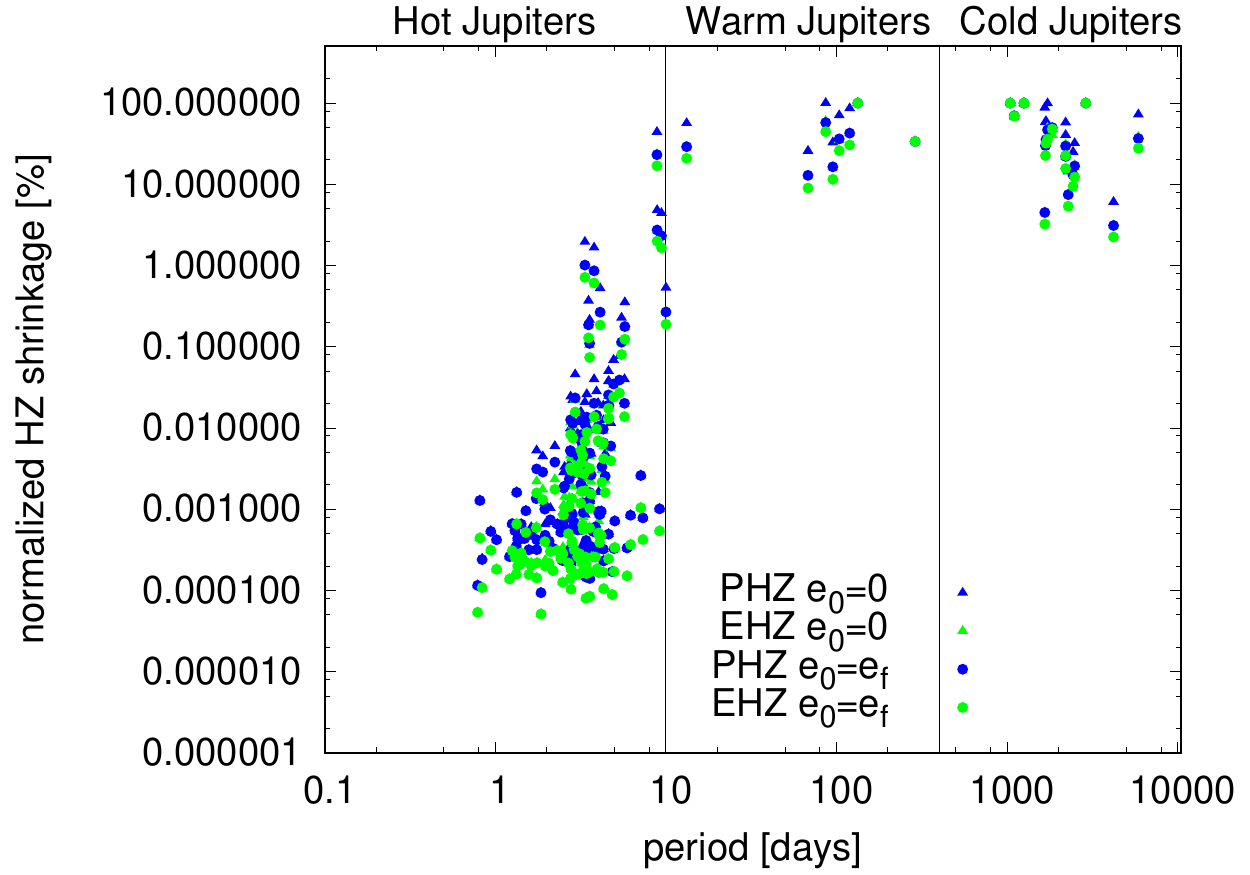}
\renewcommand{\baselinestretch}{1.}
\caption{Habitable zone shrinkages. The shrinkage of the CHZ compared to the PHZ and the EHZ is plotted against the giant planet's orbital period for $e_{0}=0$ and $e_{0}=e_f$.
\label{fig:shr2}}
\end{center}
\end{figure}

\subsection{Analytical versus numerical habitable zones}

In order to make sure that our analytical method works well, we performed a number of numerical experiments in which we 
calculated insolation extrema numerically.  For that purpose, we chose the six systems from Table 1 that had a maximum 
PHZ shrinkage of $\sim 10\%$, which makes these systems from Table 1 the most interesting candidates for that kind of numerical 
experiment.  We  also considered dynamically excited systems  (i.e. $e_0=0$) as this way any problem with the theory becomes
more noticeable.

For our simulations we used a symplectic integrator designed to integrate hierarchical triple systems (Mikkola 1997).
The code uses standard Jacobi coordinates, i.e., it calculates the relative
position and velocity vectors of the inner and outer orbit at every time step. Those were used to generate orbital elements
of the two planetary orbits and determine subsequently the borders of the PHZ.  For this series of simulations we only used the nominal values of the six systems.  

The results of the numerical simulations and the comparison against the analytical values can be found in Table 2.  As seen there, the 
analytical estimates are in excellent agreement  with the results from the numerical simulations.  We would like to point out that for the four P-type systems 
we only took into account Newtonian gravity between point masses.  This, however,  does not affect qualitatively the outcome of the comparison.
As seen in the appendix, our analytical method takes into consideration all dynamical effects that may be important for the determination of the dynamically informed habitable zones. 

\begin{table}[H]
\caption{The borders of the PHZ for six systems calculated by theoretical and numerical means. The first two columns are the borders of the classical habitable zone, while the exponent `n' denotes the values obtained from numerical simulations.} 
\vspace{0.1 cm}
\begin{center}	
{\begin{tabular}{c c c c c c c}\hline
Name  &  $ HZ_I [\text{au}]  $ & $ HZ_O [\text{au}]  $  &   $ PHZ^n_I [\text{au}]  $ & $ PHZ^n_O [\text{au}]  $  &$ PHZ_I [\text{au}]  $ & $ PHZ_O [\text{au}]  $\\
\hline\\
BD-114672  & 0.32 & 0.59 &  0.32 & 0.57 & 0.32 & 0.57\\
CoRoT-6 & 1.06 & 1.85 &  1.08 & 1.83 & 1.08 & 1.83\\
HAT-P-54  & 0.36 & 0.68 &  0.37 & 0.67 & 0.37 & 0.67\\
HD-13931  & 1.19 & 2.09 &  1.20 & 2.05 & 1.20 & 2.05\\
Kepler-63 & 0.81 & 1.43 &  0.89 & 1.35 & 0.89 & 1.35\\
WASP-89  & 0.68 & 1.23 &  0.70 & 1.21 & 0.70 & 1.21\\
\hline
\end{tabular}}
\end{center}
\end{table}

\section{Summary and discussion}\label{sec:conclusion}
Are giant planets good neighbors for habitable worlds? In terms of supporting planetary habitability, 
the answer to this question mainly depends on the location of the giant planet with respect to the 
classical habitable zone, the giant planet's eccentricity and the terrestrial planet's capacity to buffer time dependent insolation variability. 
If these factors are combined in an unfavourable fashion, giant planets diminish a system's chances of hosting a habitable world. 
This is the case for the majority of warm and cold Jupiter systems discovered so far.
The presence of close-in hot Jupiters as well as cold Jupiters farther out can be more benign as far as their influence on the habitable zone is concerned. 
Some even allow for an extension of the classical habitable zone if the terrestrial planet's climate inertia is high.
Especially when potentially habitable worlds have low climate inertia, however, giant planets have to have near-circular orbits in order not to 
reduce the size of the habitable zone. The tidal interaction with the host star practically guarantees that hot Jupiter systems fulfill such criteria. 
If terrestrial planets survive the formation and migration phase, hot Jupiters make excellent neighbors for habitable worlds.
This hypothesis is backed by our analysis of 147 exoplanet systems hosting a single gas giant orbiting a main sequence star.

In this work, we have developed a general method that can assess the level at which planetary systems  can sustain habitable conditions on Earth-like planets as we know them and hence provide observers with a tool to select possible targets in search for habitable worlds. The method has been constructed in such a way that can be applied to different planetary masses and is independent of the dynamical model and the insolation limits.  
As seen to some extent in the introduction,  our assessment of habitablity is not related to the  formation history of the system as our intention is not  to  investigate whether a terrestrial planet can form in a specific system.   The method developed here can be applied without any further assumption regarding the formation history of the system under investigation.
It is currently thought that the orbital distribution of giant planets has been mainly sculpted by two processes: orbital migration and planet-planet scattering (e.g Morbidelli 2014; Kley 2017). Each of these processes may have a strong effect on the formation and stability of rocky planets. Migration can be good or bad for terrestrial planet formation depending on the assumptions made (e.g Fogg {\&} Nelson 2005; Raymond et al. 2006; Mandell et al. 2007; Fogg {\&} Nelson 2009). The large eccentricities of many gas giants are likely to be the result of planet-planet scattering events which are decisive for the formation of rocky planets and their subsequent fate (e.g. Raymond et al. 2011, 2012; Carrera et al. 2016).
In addition, water delivery to rocky planets may also be related to the masses and eccentricities of the giant planets of the system (e.g. Raymond et al. 2004; Bancelin et al. 2016).  Rocky planets can also be formed having a significant amount of water (e.g. Raymond 2006). To conclude, it is possible that the present architecture of a planetary system may be strongly related to its formation history and subsequent dynamical evolution.  This may place additional constrains on whether a rocky planet exists in such a system.  However,  the outcome of terrestrial planet formation strongly depends on the initial 
setup and physics incorporated in the simulation and a quantitative link between results from  planet formation simulations and 
observed system configurations has not yet been established.

\acknowledgments
This research has received funding from the Jet Propulsion Laboratory through the California Institute of Technology postdoctoral fellowship program, under a contract with the National Aeronautics and Space Administration, USA,
as well as the IMCCE Observatoire de Paris, France.
The authors have used the NASA Exoplanet Archive, which is operated by the California Institute of Technology, under contract with the National Aeronautics and Space Administration under the Exoplanet Exploration Program. Furthermore, the authors would like to thank the High Performance Computing Resources team at New York University Abu Dhabi and especially Jorge Naranjo for helping us with certain aspects of the evaluation of the habitable zones.\\ 

\appendix
\label{sec:apa}
\section{Dynamical evolution}
\subsection{Orbital stability}
An important aspect of trying to determine the limits of dynamically informed habitable zones is to ensure that the terrestrial planetary orbit is stable under the gravitational perturbations of the giant planet.  In the context of this work, a system is classified as stable if the terrestrial planet
neither collides with the star nor escapes the system. Several stability criteria have been developed over the past years (Georgakarakos 2008).  Here, we make use of a criterion that applies specifically to systems with two planets (Petrovich 2015). A two planet system is stable against either ejection or collision with the star when:
\begin{eqnarray}
\frac{a_{g0}(1-e_{g0})}{a_{0}(1+e_{0})}&>&2.4\left(\frac{m_{g}}{m_*}\right)^{\frac{1}{3}}\left(\frac{a_{g0}}{a_{0}}\right)^\frac{1}{2}+1.15 \hspace{0.1cm}\mbox{(the giant planet is the outer planet)}\nonumber\\
\frac{a_{0}(1-e_{0})}{a_{g0}(1+e_{g0})}&>&2.4\left(\frac{m_{g}}{m_*}\right)^{\frac{1}{3}}\left(\frac{a_{0}}{a_{g0}}\right)^\frac{1}{2}+1.15 \hspace{0.1cm}\mbox{(the giant planet is the inner planet),}\nonumber
\end{eqnarray}
where the index 0 indicates that initial conditions rather than evolved sets of orbital elements are to be used in the above equations. 

\subsection{Orbital evolution of the terrestrial planet}
\label{sec:orbitevolution}
In order to calculate dynamically informed habitable zones, i.e., PHZ, EHZ and AHZ, we need expressions for the maximum and the averaged square eccentricity of the terrestrial planet. 
For this purpose, we use a recently published work which deals with the orbital evolution of a terrestrial planet under the gravitational influence of a giant planet (Georgakarakos et al. 2016).   This work deals with secular evolution only.
Short period effects on the orbital evolution of the terrestrial planet are more prominent when the giant planet is on a circular or nearly circular orbit.
They have been neglected in this work, since their influence on the extent of the dynamically informed HZs is believed to be small. Further investigations in this respect are necessary
to confirm this notion, however. The dynamical model describing the orbital motion of the planets is based on the assumption that all orbits are coplanar and non-resonant.  Moreover, the orbit of the giant planet is assumed to be unchanged due to the gravitational perturbations of the terrestrial planet. 
Finally, we would like to point out that  
since we deal with secular motion only, the semi-major axes of both planets are assumed to remain constant (Harrington 1968).  Given those assumptions, the components of the Earth-like planet's eccentricity vector read (Georgakarakos et al. 2016)
\begin{eqnarray}
\label{inner1}
e_{x} &=& C_1\cos{\sqrt{B^2-2AB}t}+C_2\sin{\sqrt{B^2-2AB}t}-\frac{C}{B}\cos{\varpi_g}\nonumber\\
\label{inner2}
e_{y}&=&\frac{1}{B-2A\cos^2{\varpi_g}}[(C_2\sqrt{B^2-2AB}-AC_1\sin{2\varpi_{g}})\cos{\sqrt{B^2-2AB}t}-(C_1\sqrt{B^2-2AB}+\nonumber\\& & +AC_2\sin{2\varpi_{g}})\sin{\sqrt{B^2-2AB}t}]-\frac{C}{B}\sin{\varpi_g},\nonumber
\end{eqnarray}
where $C_1$ and $C_2$ are constants of integrations. It follows that
\begin{equation}
e=\sqrt{e_{x}^2+e_{y}^2}\hspace{0.5cm}\mbox{and}\hspace{0.5cm}\varpi=\arctan{(\frac{e_{y}}{e_{x}})}.\nonumber
\end{equation}
The quantities A, B and C depend on the masses and orbital elements of the star and the two planets and they are given by equations (32)-(48) for the S-type case and equations (69)-(85) for the P-type case in the above mentioned paper. 
The required  expressions for the averaged square and maximum eccentricity of the terrestrial planet are:
\begin{equation}
\langle e^2\rangle=\left(\frac{C}{B}\right)^2\frac{2B-3A}{B-2A} \hspace{1cm}\nonumber
\end{equation}
and 
\begin{equation}
\label{emax222}
e^{max}=-2\frac{C}{B}.
\end{equation}
Also, the period of the oscillation is 
\begin{eqnarray}
P_s=\frac{2\pi}{\nu}=\frac{2\pi}{\sqrt{B^2-2AB}}.\nonumber
\end{eqnarray}
The above equations for the averaged square and maximum eccentricity hold for an initially circular terrestrial planetary orbit.  
When the initial eccentricity is equal to the forced eccentricity $e_f=-C/B$, then we have
\begin{equation}
\label{eq:ave0}
\langle e^2\rangle=\left(\frac{C}{B}\right)^2\nonumber
\end{equation}
and 
\begin{equation}
\label{emax111}
e^{max}=-\frac{C}{B}.
\end{equation}
The above expressions are valid for both S-type and P-type configurations. 

Assuming that $a/a_g \ll1$, for $e_{0}=0$, we obtain the classical result for S-type systems  (Heppenheimer 1978;
Georgakarakos 2003)
\begin{eqnarray}
\label{eq:emaxse0}
e^{max}\approx\frac{5}{2}\frac{a}{a_g}\frac{e_g}{1-e_g^2}=\epsilon, \qquad \langle e^2\rangle = \frac{1}{2}\epsilon^2.
\end{eqnarray}
For S-type systems with $e_{0}=e_f$ we find
\begin{eqnarray}
\label{eq:emaxsef}
 e^{max}\approx\frac{1}{2}\epsilon, \qquad {\langle e^2\rangle} = \frac{1}{4}\epsilon^2.
\end{eqnarray}
Similarly, for P-type configurations, assuming $a_g/a \ll1$ the corresponding equations for $e_{0}=0$ read (Moriwaki {\&}
Nakagawa 2004; Georgakarakos {\&} Eggl 2015)
\begin{eqnarray}
\label{eq:emaxpe0}
e^{max}\approx\frac{5}{4}\frac{a_g}{a}\frac{4 e_g+3 e_g^3}{2+3e_g}=\eta, \qquad \langle e^2\rangle = \frac{1}{2}\eta^2,
\end{eqnarray}
while for $e_{0}=e_f$ we get
\begin{eqnarray}
\label{eq:emaxpef}
e^{max}\approx\frac{1}{2}\eta, \qquad {\langle e^2\rangle} = \frac{1}{4}\eta^2.
\end{eqnarray}
Note that the above equations represent but the crudest approximation to the secular behavior of the two planetary orbits. The reader is advised
to follow the authors in using the full expressions (Georgakarakos et al. 2016) in order to calculate the maximum and average squared eccentricities. 
This avoids unnecessary inaccuracies in the dynamic model when the giant planet orbits in the vicinity of the habitable zone.   

\subsection{Other dynamical effects}
When the giant planet is close to the star, other dynamical effects than Newtonian gravity between point masses become important. 
The main contribution of these effects, namely general relativity, non-dissipative tidal bulge deformation and deformation due to rotation, is an increase in the precession rate of the pericenter of the giant planet. 
The relevant precession rates,  assuming alignment between the spin axis and the orbit normal, are (e.g.
Fabrycky {\&} Tremaine 2007; Fabrycky 2010):
\begin{equation}
\dot{\varpi}_{gr}=\frac{3G^{\frac{3}{2}}(m_*+m_{g})^{\frac{3}{2}}}{a^{\frac{5}{2}}_{g}c^2(1-e^2_{g})},\nonumber
\end{equation}
\begin{equation}
\dot{\varpi}_{tb}=\frac{15\sqrt{G(m_*+m_{g})}}{8a^{\frac{13}{2}}_{g}}\frac{8+12e^2_{g}+e^4_{g}}{(1-e^2_{g})^5}\left(\frac{m_{g}}{m_*}k_{*}R^5_*+\frac{m_{*}}{m_{g}}k_{g}R^5_{g}\right)\nonumber
\end{equation}
\begin{equation}
\dot{\varpi}_{rot}=\frac{\sqrt{m_*+m_{g}}}{\sqrt{G}a^{\frac{7}{2}}_{g}(1-e^2_{g})^2}\left(\frac{k_*R^5_*}{m_*}\Omega_*+\frac{k_{g}R^5_{g}}{m_{g}}\Omega_{g}\right)\nonumber,
\end{equation}
where $G$ is the gravitational constant, $c$ is the speed of light, $R_*$ denotes the radius of the star, $R_{g}$ is the radius of the giant planet, $k_*$ and $k_{g}$ denote the classical apsidal motion constants of the star and the giant planet respectively (Russell 1928), $\Omega_*$ denotes the stellar spin angular velocity and $\Omega_{g}$ is the giant planetary spin angular velocity.  For the latter, we use the following pseudo-synchronization expression (Hut 1981)
\begin{equation}
\Omega_{g}=n_{g}\frac{1+(15/2)e^2_{g}+(45/8)e^4_{g}+(5/16)e^6_{g}}{[1+3e^2_{g}+(3/8)e^4_{g}](1-e^2_{g})^{\frac{3}{2}}},\nonumber
\end{equation}
where $n_{g}$ is the mean motion of the giant planet.
For the apsidal motion constants we use the values $k_*=0.014$ and $k_{g}=0.25$ (Fabrycky {\&} Tremaine 2007).

The analytical model that describes the orbital evolution of the terrestrial planet (Georgakarakos et al. 2016) assumes that the pericenter of the giant planet remains constant.  Therefore, in order to incorporate the above dynamical effects in our model, we use an orbital solution that allows the inclusion of such effects and has been developed originally to model circumbinary planetary motion (Georgakarakos {\&} Eggl 2015).  In addition, 
that orbital solution includes short periodic terms. When the giant planet is on a nearly circular orbit ($e_{g}\sim0.001$ or less) the orbit of the perturbed body
shows no secular evolution and consequently, in the absence of a mean motion resonance, the motion is dominated by short period effects.
Hence, for P-type systems that require the addition of the 
previously mentioned dynamical effects, the relevant eccentricity expressions are the following:
\begin{eqnarray}
\label{ave1}
\langle e^2\rangle&=&\frac{m^2_{*}m^2_{g}}{(m_{*}+m_{g})^{\frac{8}{3}}M^{\frac{4}{3}}}\frac{1}{X^{\frac{8}{3}}}
\bigg[\frac{9}{8}+\frac{27}{8}e^2_{g}+\frac{887}{64}e^4_{g}-\frac{975}{64}\frac{1}{X}e^4_{g}\sqrt{1-e^2_{g}}+\frac{1}{X^2}\Big(\frac{225}{64}+\frac{6619}{64}e^2_{g}-\nonumber\\
& & -\frac{26309}{512}e^4_{g}-\frac{393}{64}e^6_{g}\Big)\bigg]+2\Big(\frac{K_{2}}{K_1-K_3}\Big)^2.
\end{eqnarray}
\begin{equation}
\label{max1}
e^{max}=\frac{m_{*}m_{g}}{(m_{*}+m_{g})^{\frac{4}{3}}M^{\frac{2}{3}}}\frac{1}{X^{\frac{4}{3}}}\bigg[\frac{3}{2}+
\frac{17}{2}e^2_{g}+\frac{1}{X}\Big(3+19e_{g}+\frac{21}{8}e^2_{g}-\frac{3}{2}e^3_{g}\Big)\bigg]+2\left|\frac{K_2}{K_1-K_3}\right|.
\end{equation}
where
\begin{eqnarray}
K_1 & = \frac{3}{8}\frac{\sqrt{\mathcal{G}M}m_*m_{g}a^2_{g}}{(m_*+m_{g})^2a^{\frac{7}{2}}}(2+3e^2_{g})\nonumber\\
K_2 & = \frac{15}{64}\frac{\sqrt{\mathcal{G}M}m_*m_{g}(m_*-m_{g})a^3_{g}}{(m_*+m_{g})^3a^{\frac{9}{2}}}e_{g}(4+3e^2_{g})\nonumber\\
K_3 & = \dot{\varpi}_{gr}+\dot{\varpi}_{hb}+\dot{\varpi}_{rot}. \label{eq:K}\nonumber
\end{eqnarray}
$M$ is the total mass of the system and $X$ is the ratio of the the two planetary orbital periods given by
\begin{equation}
X=\sqrt{\frac{m_*+m_{g}}{M}\left(\frac{a}{a_{g}}\right)^3}.\nonumber
\end{equation}
Equations (\ref{ave1}) and (\ref{max1}) apply to an initially circular terrestrial planetary orbit. If the initial eccentricity is equal to the forced eccentricity 
then the factor 2 in front of the $K$ factors (which denote the secular contribution) should be dropped.

\subsection{Explicit formulae for the PHZ and AHZ}\label{sec:explicit}
Although the formulae for the dynamically informed HZ borders have to be calculated numerically in most cases, one can acquire an explicit analytical solution for the PHZ and the AHZ borders if a simple secular orbit evolution model is implemented.
If one uses a first order secular evolution model for the
planetary orbit, for instance, by combining equations (\ref{eq:phzio}), (\ref{eq:emaxse0}) and (\ref{eq:emaxsef})
we find the following PHZ limits for S-type systems:
\begin{equation}
\label{eq:phzsti}
\quad a_{I}=\frac{p}{5\; w\; e_g}\left[1-\left(1-\frac{10\; w\; e_g}{p}\;r_{I}  \right)^{1/2} \right]\hspace{0.6cm}\mbox{and}\hspace{0.3cm}\quad a_{O}=\frac{p}{5\; w\; e_g}\left[-1+\left(1+\frac{10\; w\; e_g}{p}\;r_{O}  \right)^{1/2} \right],
\end{equation}
where $a_{I}$ and $a_{O}$ are the inner and outer borders of the PHZ and $p=a_g\left(1-e_g^2\right)$ is the semilatus rectum of the giant planet's orbit. Furthermore, $w$ is a weighting factor that depends on the initial eccentricity of the terrestrial planet's orbit and $w=1$, if $e_{0}=0$, while $w=1/2$ for $e_{0}=e_f$.
It may seem that equations (\ref{eq:phzsti}) for the inner and outer PHZ border become singular for $e_g\rightarrow0$. This is not the case, however, as the limit reads
\begin{equation}
\lim_{e_g\rightarrow0}\;a_{I,O} =r_{I,O}.\nonumber
\end{equation}
In other words, for a giant planet on a circular orbit, the PHZ becomes identical to the classical habitable zone. 
Similarly, for P-type systems and using equations (4), (A5) and (A6), the corresponding PHZ equations read
\begin{equation}
\label{eq:phzpt}
\quad a_{I}=r_{I}+ \frac{5}{4}\frac{w\; a_g\; e_g\; (4 + 3\; e_g^2)}{2+3\; e_g}\qquad\mbox{and}\hspace{0.5cm} a_{O}=r_{O}- \frac{5}{4}\frac{w\; a_g\; e_g\; (4 + 3\; e_g^2)}{2+3\;e_g},\nonumber
\end{equation}
with $w$ having the same values as previously.

An explicit solution for the AHZ can be obtained if one uses equations (5), (A3), (A4), (A5) and (A6). For S-type systems we get: 
\begin{equation}
\label{eq:ahzst}
\quad a_{I,O}=2^{3/2}\; p\; r_{I,O}\; \left[8\; p^2 - 25\;w^2\; e_g^2\; \left(r_{I,O}\right)^2  \right]^{-1/2}
\end{equation}
with $w=1/\sqrt 2$ for $e_{0}=0$, and $w=1/2$ for $e_{0}=e_f$.  
For a P-type configuration, we obtain:
\begin{equation}
\label{eq:ahzpt}
\text{AHZ:} \quad a_{I,O}=\frac{r_{I,O}}{\sqrt{2}}\left\{1+\left[1 + \frac{25}{8}\;w^2 a_g^2 \left(r_{I,O}\right)^{-2} \; \left(\frac{4\; e_g +3\; e_g^3}{2+3\; e_g}\right)^2 \right]^{1/2}\right\}^{1/2},
\end{equation}
where $w$ has the same value as in the S-type case.
The equations governing the EHZ do not allow for trivial solutions even using simplified secular dynamics.

\subsection{Mass and rotational velocity calculation}
For systems where we only know $m_{g}\sin{i}$ ($i$ being the inclination of the planet's orbit to the plane perpendicular to the line-of-sight) 
and not the real mass of the giant planet we use the expected value of the mass. The same approach is applied to stellar rotational velocities $V_*\sin{i}$. 
\label{sec:apc}
We may write $m_{g}=m_{g}\sin{i}/\sin{i}=m_{sg}/\sin{i}$, where $i$ is the inclination 
angle between the orbital plane and the plane of the sky.  Adopting the $13M_J$ limit as the
upper bound for the giant planet's mass, this is achieved when the angle $i$ is $i_b=\arcsin{(m_{sg}}/13)$.  Assuming that the probability to have a specific angle $i$ is the same for all $i$, we get for the expected value of $m_{g}$:
\begin{equation}
<m_{g}>=\frac{1}{\pi/2-i_b}\int_{i_b}^{\frac{\pi}{2}}\frac{m_{sg}}{\sin{i}}di=
\frac{m_{sg}}{\pi/2-i_b}|-\ln{(\tan{\frac{i_b}{2}})}|.\nonumber
\end{equation}
Similarly, we can apply the same idea for the stellar rotational velocity.  For the upper limit of the velocity we adopt the equatorial 
break-up velocity (Maeder {\&} Meynet 2000) $V_{crit}=(2Gm_*/3R_*)^{1/2}$. If we denote $V_{s*}=V_{*}\sin{i}$ and 
$i_b=\arcsin{(V_{s*}}/V_{crit})$, then we have:
\begin{equation}
<V_{*}>=\frac{1}{\pi/2-i_b}\int_{i_b}^{\frac{\pi}{2}}\frac{V_{s*}}{\sin{i}}di=
\frac{V_{s*}}{\pi/2-i_b}|-\ln{(\tan{\frac{i_b}{2}})}|.\nonumber
\end{equation}
The masses used in this work as well as the rotational velocities can be found in Appendix B.\\

\section{Data Tables}
Table B1 provides the orbital elements and the physical parameters of our systems. 
The corresponding data was extracted from the NASA Exoplanet Archive (http://exoplanetarchive.ipac.caltech.edu). 
Table B2 gives the borders of all habitable zones for systems with the terrestrial planet having formed on an initially circular orbit $e_0=0$. Table B3 contains the habitable zone shrinkages for systems with $e_0=0$. 
Tables B4 and B5 are similar to Tables B2 and B3 but for systems with $e_0=e_f$, i.e. where the terrestrial planet had an initially eccentric orbit.
The occurrence of the `*' symbol in Tables B2 and B4 indicates that the perturbations of the giant planet are strong enough to 
render the respective system uninhabitable (i.e. the respective habitable zone vanishes).
Tables B6 and B7 give the expected values of the giant planet masses and the stellar rotational velocities respectively. All the values in our tables are given with their uncertainties.   Finally, Table B8 contains a list of the symbols, variables and acronyms
used in this work.

\newpage
\label{sec:apb}
{\footnotesize
\small
\setlength\tabcolsep{3pt}


\begin{table}[H]
\caption{Expected values for giant planet masses and their uncertainties.\label{tab:mav}} 
\vspace{0.1 cm}
\begin{center}	
{\begin{tabular}{c c c c}\hline
Name  &  Giant planet mass & Name  &  Giant planet mass\\
\hline\\
BD-114672  &$ 1.348^{+ 11.652}_{ -0.868}$ &HD25171    &$ 2.098^{+ 10.902}_{ -1.248}$\\
GJ179      &$ 1.879^{+ 11.121}_{ -1.129}$ &HD27631    &$2.866^{+ 10.134}_{ -1.556}$ \\
GJ3021     &$5.217^{+ 7.783}_{ -1.937}$  &HD290327   &$4.281^{+ 8.719}_{ -1.881}$\\
HD109246   &$ 1.792^{+ 11.208}_{ -1.112}$ &HD30669   &$ 1.229^{+ 11.771}_{ -0.819}$\\
HD13931    &$3.457^{+ 9.543}_{ -1.727}$  &HD52265    &$2.261^{+ 10.739}_{ -1.241}$\\
HD143361   &$5.335^{+ 7.665}_{ -2.095}$  &HD63454    &$ 1.080^{11.920}_{-0.682}$ \\
HD145377   &$7.553^{+ 5.467}_{ -1.873}$  &HD6718     &$3.022^{+ 9.978}_{ -1.562}$\\
HD150706   &$4.481^{+8.519}_{ -2.431}$   &HD68402    &$4.890^{+ 8.110}_{ -2.170}$ \\
HD152079   &$3.842^{+ 9.158}_{ -1.832}$   &HD73267    &$4.879^{+ 8.121}_{ -1.889}$\\
HD204941   &$ 0.786^{+ 12.214}_{ -5.552}$ &HD86226 &$ 2.049^{+ 10.951}_{ -1.229}$\\
HD220689   &$2.276^{+ 10.724}_{ -1.306}$\\
\hline
\end{tabular}}
\end{center}
\end{table}

\begin{longtable}{l l l l l l l l}
\caption{Expected values for stellar rotation velocities and their uncertainties.\label{tab:vav}} 
\vspace{0.1 cm}\\
\hline\\
Name  &  Velocity & Name  &  Velocity & Name  &  Velocity & Name  &  Velocity\\
\hline\\
BD-114672  &$ 5.651^{+73.123}_{ -3.651}$&HAT-P-54   &$ 6.382^{+70.657}_{ -4.532}$&HD86226    &$ 6.481^{+70.220}_{ -4.081}$&WASP-23    &$ 6.038^{+69.689}_{ -4.138}$\\
CoRoT-10   &$ 5.667^{+74.193}_{ -4.167}$&HAT-P-55   &$ 5.150^{+70.090}_{ -3.850}$&K2-29      &$ 9.104^{+69.557}_{ -5.904}$&WASP-25    &$ 7.745^{+70.698}_{ -5.745}$\\
CoRoT-12   &$ 3.208^{+70.676}_{ -3.208}$&HATS-1     &$ 6.667^{+66.742}_{ -4.647}$&K2-30      &$ 4.248^{+73.633}_{ -3.148}$&WASP-26    &$ 9.200^{+60.325}_{ -5.700}$\\
CoRoT-13   &$ 9.647^{+68.516}_{ -6.647}$&HATS-10    &$12.443^{+62.797}_{ -7.463}$&K2-31      &$ 6.987^{+74.282}_{ -4.887}$&WASP-28    &$ 8.096^{+64.688}_{ -5.186}$\\
\end{longtable}

\begin{table}[H]
\caption{Notations} 
\begin{center}
{\begin{tabular}{c c}\hline
\renewcommand{\baselinestretch}{1.}
Habitable Zones & \\\hline
HZ & habitable zone\\
AHZ & averaged habitable zone\\
CHZ & classical habitable zone\\
DIHZ & dynamically informed habitable zone (PHZ, AHZ, EHZ)\\
EHZ & extended habitable zone\\
PHZ & permanently habitable zone\\\hline
Non-SI Units & \\
\hline
\end{tabular}}
\end{center}
\end{table}
\newpage
\begin{center}
REFERENCES
\end{center}
Agnew, M. T., Maddison, S. T., Thilliez, E., {\&} Horner, J. 2017,
MNRAS, 471, 4494\\
Arnold, V. I. 1978, Mathematical methods of classical mechanics\\
Bancelin, D., Pilat-Lohinger, E., {\&} Bazso, A. 2016, A{\&}A, 591,
A120\\
Bolmont, E., Libert, A.-S., Leconte, J., {\&} Selsis, F. 2016, A{\&}A,
591, A106\\
Carrera, D., Davies, M. B., {\&} Johansen, A. 2016, MNRAS, 463,
3226\\
Cockell, C. S., Bush, T., Bryce, C., et al. 2016, Astrobiology, 16,
89\\
Davis, M., Hut, P., {\&} Muller, R. A. 1984, Nature, 308, 715\\
Dressing, C. D., Spiegel, D. S., Scharf, C. A., Menou, K., {\&}
Raymond, S. N. 2010, ApJ, 721, 1295\\
Dvorak, R., Eggl, S., Süli, Á., et al. 2012, in American Institute
of Physics Conference Series, Vol. 1468, American Institute of
Physics Conference Series, ed. M. Robnik {\&} V. G.
Romanovski, 137–147\\
Dvorak, R., Pilat-Lohinger, E., Bois, E., et al. 2010,
Astrobiology, 10, 33\\
Eggl, S., Pilat-Lohinger, E., Funk, B., Georgakarakos, N., {\&}
Haghighipour, N. 2013, MNRAS, 428, 3104\\
Eggl, S., Pilat-Lohinger, E., Georgakarakos, N., Gyergyovits, M.,
{\&} Funk, B. 2012, ApJ, 752, 74\\
Fabrycky, D., {\&} Tremaine, S. 2007, ApJ, 669, 1298
Fabrycky, D. C. 2010, Non-Keplerian Dynamics of Exoplanets,
ed. S. Seager, 217–238\\
Fogg, M. J., {\&} Nelson, R. P. 2005, A{\&}A, 441, 791\\
Fogg, M. J., {\&} Nelson, R. P. 2007, A{\&}A, 461, 1195\\
Fogg, M. J., {\&} Nelson, R. P. 2009, A{\&}A, 498, 575\\
Georgakarakos, N. 2003, MNRAS, 345, 340\\
Georgakarakos,N. 2008, Celestial Mechanics and Dynamical Astronomy, 100,
151\\
Georgakarakos, N., Dobbs-Dixon, I., {\&} Way, M. J. 2016,
MNRAS, 461, 1512\\
Georgakarakos, N., {\&} Eggl, S. 2015, ApJ, 802, 94\\
Grazier, K. R. 2016, Astrobiology, 16, 23\\
Harrington, R. S. 1968, AJ, 73, 190\\
Heller, R., Williams, D., Kipping, D., et al. 2014, Astrobiology,
14, 798\\
Heppenheimer, T. A. 1978, A{\&}A, 65, 421\\
Horner, J., {\&} Jones, B. W. 2008, International Journal of
Astrobiology, 7, 251\\
Horner, J., {\&} Jones, B. W. 2009, International Journal of Astrobiology, 8, 75\\
Horner, J., {\&} Jones, B. W. 2011, Astronomy and Geophysics, 52, 1.16\\
Horner, J., Jones, B. W., {\&} Chambers, J. 2010, International
Journal of Astrobiology, 9, 1\\
Huang, C., Wu, Y., {\&} Triaud, A. H. M. J. 2016, ApJ, 825, 98\\
Hut, P. 1981, A{\&}A, 99, 126\\
Izidoro, A., Morbidelli, A., {\&} Raymond, S. N. 2014, ApJ, 794, 11\\
zidoro, A., Raymond, S. N., Morbidelli, A., Hersant, F., {\&}
Pierens, A. 2015, ApJL, 800, L22\\
Jones, B. W., Underwood, D. R., {\&} Sleep, P. N. 2005, ApJ, 622,
1091\\
Kaiho, K., {\&} Oshima, N. 2017, Scientific Reports, 7, 14855\\
Kasting, J. F., Whitmire, D. P., {\&} Reynolds, R. T. 1993, Icarus,
101, 108\\
Kaufmann, R. K., {\&} Juselius, K. 2016, Paleoceanography, 31, 286\\
Kley, W. 2017, ArXiv e-prints, arXiv:1707.07148\\
Kopparapu, R. K., {\&} Barnes, R. 2010, ApJ, 716, 1336\\
Kopparapu, R. K., Ramirez, R. M., SchottelKotte, J., et al.
2014, ApJL, 787, L29\\
Kopparapu, R. K., Ramirez, R., Kasting, J. F., et al. 2013, ApJ,
765, 131\\
Kring, D. A. 2003, Astrobiology, 3, 133\\
Laskar, J., Robutel, P., Joutel, F., et al. 2004, A{\&}A, 428, 261\\
Maeder, A., {\&} Meynet, G. 2000, A{\&}A, 361, 159\\
Mandell, A. M., Raymond, S. N., {\&} Sigurdsson, S. 2007, ApJ,
660, 823\\
Mardling, R. A. 2007, MNRAS, 382, 1768\\
Menou, K., {\&} Tabachnik, S. 2003, ApJ, 583, 473\\
Mikkola, S. 1997, Celestial Mechanics and Dynamical
Astronomy, 67, 145\\
Morbidelli, A. 2014, Philosophical Transactions of the Royal
Society of London Series A, 372, 20130072\\
Moriwaki, K., {\&} Nakagawa, Y. 2004, ApJ, 609, 1065\\
Petrovich, C. 2015, ApJ, 808, 120\\
Popp, M., {\&} Eggl, S. 2017, Nature Communications, 8, 14957\\
Raymond, S. N. 2006, ApJL, 643, L131\\
Raymond, S. N., Barnes, R., {\&} Kaib, N. A. 2006, ApJ, 644, 1223\\
Raymond, S. N., Quinn, T., {\&} Lunine, J. I. 2004, Icarus, 168, 1\\
Raymond, S. N., Armitage, P. J., Moro-Martı́n, A., et al. 2011,
A{\&}A, 530, A62\\
Raymond, S. N., Armitage, P. J., Moro-Martı́n, A., et al. 2012, A{\&}A, 541, A11
Russell, H. N. 1928, MNRAS, 88, 641\\
Sándor, Z., Süli, Á., Érdi, B., Pilat-Lohinger, E., {\&} Dvorak, R.
2007, MNRAS, 375, 1495\\
Schwarz, R., Pilat-Lohinger, E., Dvorak, R., Érdi, B., {\&} Sándor,
Z. 2005, Astrobiology, 5, 579\\
Spalding, C., Batygin, K., {\&} Adams, F. C. 2016, ApJ, 817, 18\\
Spiegel, D. S., Raymond, S. N., Dressing, C. D., Scharf, C. A., {\&}\\
Mitchell, J. L. 2010, ApJ, 721, 1308\\
Steffen, J. H., Ragozzine, D., Fabrycky, D. C., et al. 2012,
Proceedings of the National Academy of Science, 109, 7982\\
von Bloh, W., Bounama, C., {\&} Franck, S. 2007,
Planet. Space Sci., 55, 651\\
Way, M. J., {\&} Georgakarakos, N. 2017, ApJL, 835, L1\\
Williams, D. M., {\&} Pollard, D. 2002, International Journal of
Astrobiology, 1, 61\\
\end{document}